\documentclass[ aps,twocolumn, floatfix, showpacs]{revtex4}
\usepackage{graphicx,epsfig,amsmath,amsfonts,epic,eepic,psfrag,latexsym,
amssymb}

\begin{document}

\title{Mechanism of murine epidermal maintenance: Cell division and the Voter Model}
\author{Allon M. Klein$^1$, David P. Doup\'e$^2$, Philip H. Jones$^2$, and Benjamin D. Simons$^1$}
\affiliation{$^1$Cavendish Laboratory, Madingley Road, Cambridge CB3 OHE, UK\\
$^2$ MRC Cancer Cell Unit, Hutchison-MRC Research Centre, Cambridge CB2 2XZ, 
UK}

\textwidth = 7in


\begin{abstract}
The dynamics of a genetically-labelled 
cell population may be used to infer the laws of cell division in mammalian tissue.
Recently, we showed that in mouse tail-skin, where proliferating cells are confined 
to a two-dimensional layer, cells proliferate and differentiate according to
a simple stochastic model of cell division involving just one type
of proliferating cell that may divide both symmetrically and asymmetrically. 
Curiously, these simple rules
provide excellent predictions of the cell population dynamics
without having to address their spatial distribution.
Yet, if the spatial behaviour of cells is addressed by 
allowing cells to diffuse at random, 
one deduces that density fluctuations destroy tissue confluence,
implying some hidden degree of spatial regulation in the physical system.
To infer the mechanism of spatial regulation, 
we consider a two-dimensional model 
of cell fate that preserves the overall population dynamics. 
By identifying the resulting behaviour with a three-species variation 
of the ``Voter'' model, we predict that 
 proliferating cells in the basal layer should \emph{cluster}. 
Analysis of empirical correlations of cells stained for
proliferation activity confirms that the expected clustering behaviour is
indeed seen in nature. As well as explaining how cells maintain 
a uniform two-dimensional density, these findings present
an interesting experimental example of voter-model statistics in biology.
\end{abstract}

\maketitle

\section{Introduction}
A major challenge in biology is to determine how proliferating
cells behave in developing and adult tissues.
To gain insight into the processes of cancer onset, aging and wound healing,
biologists have long recognised that the spatial organisation of
cells in tissue provides indirect access to the underlying cell behaviour. 
Some tissues, such as the auditory hair cells of the inner ear are arranged into repeating units containing groups of specialised cells essential for the function of the tissue~\cite{Lewis:Davies:02}.  In contrast, in other tissues cells do not organize into coherent structures that reflect their cooperative function. 
%lie in a patterned array, 
Indeed, the arrangement of some cell types appears random~\cite{Braun:03}.  Inferring the rules of cell behaviour in these apparently unstructured tissues appears challenging. One may ask, therefore, how
cell behaviour in such tissues is regulated in the absence of well defined spatial
roles.

\begin{figure}
\begin{center}
\includegraphics[width=3.3in]{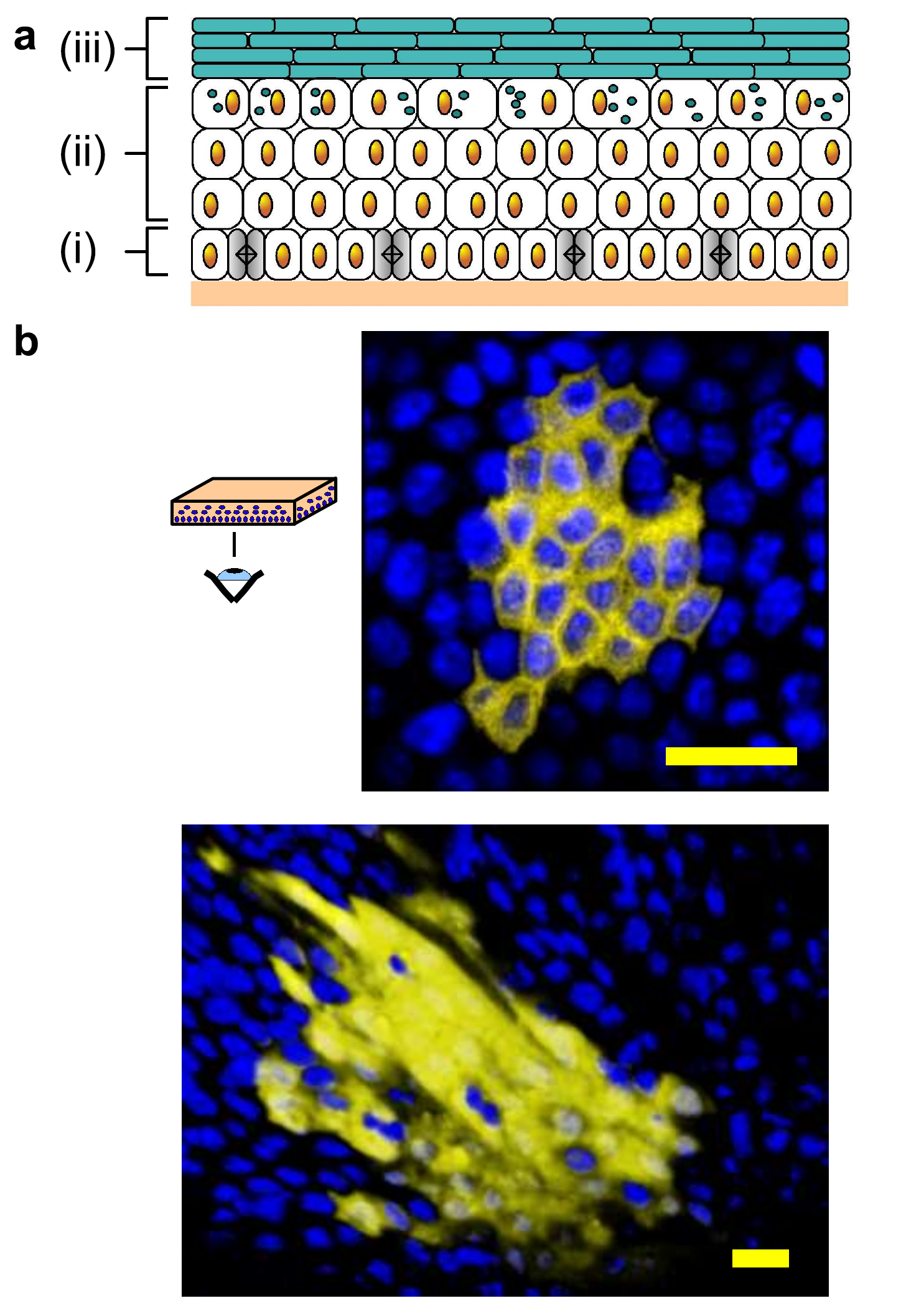}
\caption{\small 
(Color online) 
(a) Schematic cross-section of murine interfollicular epidermis (IFE). 
Proliferating cells (grey) 
are confined to the basal layer (labelled i); differentiated cells 
migrate through the superbasal layers (ii), where they flatten 
into cornified cells, losing their nuclei and assembling a 
cornified envelope (green online) (iii), eventually becoming shed 
at the surface.
(b) Two examples of typical clones acquired at a late time point, viewed from the basal layer surface. Cell nuclei are labelled blue; the 
hereditary clone marker (EYFP) appears yellow. Scale bar: $20\mu$m.
}
\label{fig:skinCrossSection}
\end{center}
\end{figure}
In this context, it is interesting to consider the case of mouse tail-skin epidermis, 
for which the laws governing cell behaviour have recently been resolved.
Mammalian epidermis comprises hair follicles interspersed with interfollicular epidermis (IFE),
which consists of sheets of specialised cells known as keratinocytes~\cite{Fuchs:07}, see
fig.~\ref{fig:skinCrossSection}(a).  Cells are shed continually from the epidermal surface, and are replaced by proliferating cells in the basal layer, whose progeny may cease proliferating and then migrate through the suprabasal layers before reaching the epidermal surface. Until recently, it was widely assumed that adult tissue is maintained by two different proliferating cell populations. These comprised long-lived, self-renewing stem cells, that have the potential to undergo an unlimited number of cell divisions and which maintain a second population of transit-amplifying cells (TA), whose proliferative potential is limited~\cite{Potten:74}.  After several rounds of division, it was conjectured that TA cells differentiate, exit the cell cycle and move out of the basal layer. 

In a recent study by Clayton \emph{et al.}, inducible genetic labelling was used to study the mechanism of epidermal maintenance by tracking the fate of a representative sample of
cells and their progeny (clones) in normal murine tail epidermis~\cite{Clayton:07}. 
By analysing the size distribution of such clones over a 
period of one year, it was possible to infer that the epidermis is maintained by 
%identify a novel and remarkably simple model of epidermal maintenance involving 
just one type of progenitor cell.  
According to the revised model of epidermal maintenance,
%introduced by Clayton \emph{et al.}, 
progenitor cells capable of both symmetric and 
asymmetric division give rise to a population of non-proliferating cells, 
which then transfer from the basal layer to the suprabasal layers. More
precisely, labelling proliferating cells as type A and differentiated basal 
layer cells as type B, the basal layer cell population is governed by the 
stochastic non-equilibrium (Markovian) process,
\begin{equation}
\label{eqn:model_0d}
\begin{array}{lcl}
&&{\rm A}\stackrel{\lambda}{\longrightarrow}
        \left\{\begin{array}{cl}
        {\rm A}+{\rm A} & {\rm Prob.\ }r \\
        {\rm A}+{\rm B} &  {\rm Prob.\ }1-2r\\
        {\rm B}+{\rm B} &  {\rm Prob.\ }r 
        \end{array}\right. \\
&&{\rm B}\stackrel{\Gamma}{\longrightarrow}\oslash\,,\\
\end{array}
\end{equation}
involving three adjustable parameters: the overall cell division rate, 
$\lambda$, the proportion of cell divisions that are symmetric, $2\times r$, 
and the rate of transfer, $\Gamma$, of non-proliferating cells from the basal 
to the suprabasal layers. 
As well as overturning the accepted paradigm
of epidermal homeostasis being achieved by discrete populations of stem and TA cells, 
the model provides a degree of quantitative predictive rigour that is unusual in the field of cell tissue biology; for example by quantifying the division and migration rates ($\lambda=1.1$/week, $\Gamma=0.31$/week) and the branching ratio ($r=0.08$), as shown in Refs.~\cite{Clayton:07, Klein:07}. 
%REVISION
As a result of its quantitative nature, the model establishes a platform for investigating the role of different cellular constituents (such as gene products) in regulating cell fate, for example by studying how different genes influence the model parameters.

It is an interesting fact that process (\ref{eqn:model_0d}) is capable of fitting the wide range of clone fate data within a ``zero-dimensional'' framework, i.e. without having to
address the spatial orientation of cells within the basal layer. 
%That such analysis is at all possible is due, in part, to the \emph{cohesive} nature of labelled clones (see for example Fig.~\ref{fig:cloneExample}), which allows unambiguous identification of distinct cell lineages. This leads one to postulate whether the role of cell mobility may be accounted for within the
Yet, the observed uniformity of cell density implies a degree of regulation beyond that which can be addressed in the zero dimensional framework.
In particular, when augmented by spatial diffusion, the proposed model leads to 
``cluster'' formation in the two-dimensional system whereupon local cell 
densities are predicted to diverge logarithmically~\cite{Houchmandzadeh:02,Young:01}. 
%REVISION
In biological terms, such behaviour would correspond to a severe disruption of the epidermis, with much of the tissue dying away, leaving only a few isolated and very thick clusters of epidermis.
Significantly, the observation that labelled families of cells remain largely \emph{cohesive} (see for example Fig.~\ref{fig:skinCrossSection}b), reveals that cell mobility must be small, so that such divergences would be significant within a mammalian lifetime. These divergences can not be regulated through a local density-dependent mobility.
%
%Yet, it is an interesting fact that such clonal analysis is at all possible because clones appear to remain cohesive over the entire period of observation. This is to some extent surprising, given that basal layer cells have the capacity to migrate within the basal layer. One may therefore ask whether it is possible to account for this characteristic spatial behaviour of basal layer cells? 
%
%

Thus, the success of the zero-dimensional fit, despite the predicted divergence in two dimensions, leaves us with an interesting challenge that is the focus of the present study: Can we uncover, from the spatial distribution of basal layer cells, the mechanism by which cells regulate a uniform cell density without compromising the integrity of the zero-dimensional fit, as embodied in process (\ref{eqn:model_0d})? 
%In particular, any mechanism regulating cell behaviour must result, in the steady-state, in an effectively non-interacting cell behaviour. Although a range of such models may exist, the ready availability of spatial data in the form of basal layer micrographs presents a proving ground on which to refine and select the possible candidates. 

In order to identify the underlying rules of cell division and differentiation, we shall draw upon the results of two types of experiment. First, we shall revisit the clone fate data used by Clayton \emph{et al.}, and examine the previously-discarded spatial distribution of labelled basal layer cells for signatures of underlying regulation. Second, we shall consider the statistics of the entire population of basal layer cells. In particular, by immunostaining basal layer cells for markers of cell proliferation, it is possible to analyse the spatial distribution of all progenitor cells. 

Thus, the aim of this paper is to elucidate how the experimental observations constrain any proposed theory of spatial behaviour in the basal layer. In summary, 
we shall show that the dynamics predicted by process (\ref{eqn:model_0d}) are indeed consistent with the constraint on uniform cell density, provided that cell division occurs only upon the migration of a nearby type B (i.e. differentiated) cell into the basal layer. Moreover, we confirm that the clone fate data is consistent with a restricted degree of cell mobility, whereby cell motion is \emph{not} diffusive and random. Instead, differentiated cells only migrate laterally as a response to fluctuations in the local density. Finally, to test the validity of the proposed spatial model, we use it to predict that, 
%REVISION
while maintaining a uniform \emph{total} areal cell density, the population of 
progenitor cells should \emph{cluster} over time. By considering the radial correlation function for the spatial distribution of progenitor cells, we find that this prediction is
in good qualitative agreement with experiment. Quantitatively, the comparison reveals that the experimental degree of progenitor cell clustering is slightly higher than that expected for the parameter value of $r=0.08$ determined previously through clonal analysis. Although several technical difficulties may challenge the reliability of these quantitative results, we speculate that such excess clustering may be a signature of spatial regulation of cell fate during asymmetric division. These results also shed light on previous observations of clustering of cells undergoing mitosis in the epidermis~\cite{Gibbs:70, Cameron:66}. Such observations have been interpreted in the biological community to be a signature of regulation that leads to coordinated cell division. By contrast, this work shows that the tendency of proliferating cells (and therefore mitoses) to cluster is in fact consistent with cells dividing independently and stochastically --- indeed it is the hallmark of the proposed spatial process.

This paper is organised as follows. 
In section~\ref{sec:latticeModel} we develop a phenomenological model of cell behaviour that incorporates the experimental constraint on uniform cell density.
% by drawing upon the recent literature on non-equilibrium lattice models. 
We identify the proposed model as a variation upon the so-called \emph{monomer-monomer} model of surface catalysis. We then analyse this simpler model in section~\ref{sec:voter}, including an exact solution for the two-point density correlations in the closely-related monomer-monomer model. 
In section~\ref{sec:experiment} we test the uniform-density model against a range of experimental data, first through a qualitative comparison of the model with the empirical clone shape data (section~\ref{sec:clone}), and then by analysing the spatial correlations observed for proliferating cells (section~\ref{sec:corr}). We conclude with a discussion of cell clustering in section~\ref{sec:disc}.

\section{A spatial model of cell kinetics}
\label{sec:model}
\begin{figure}
\begin{center}
\includegraphics[width=3.3in]{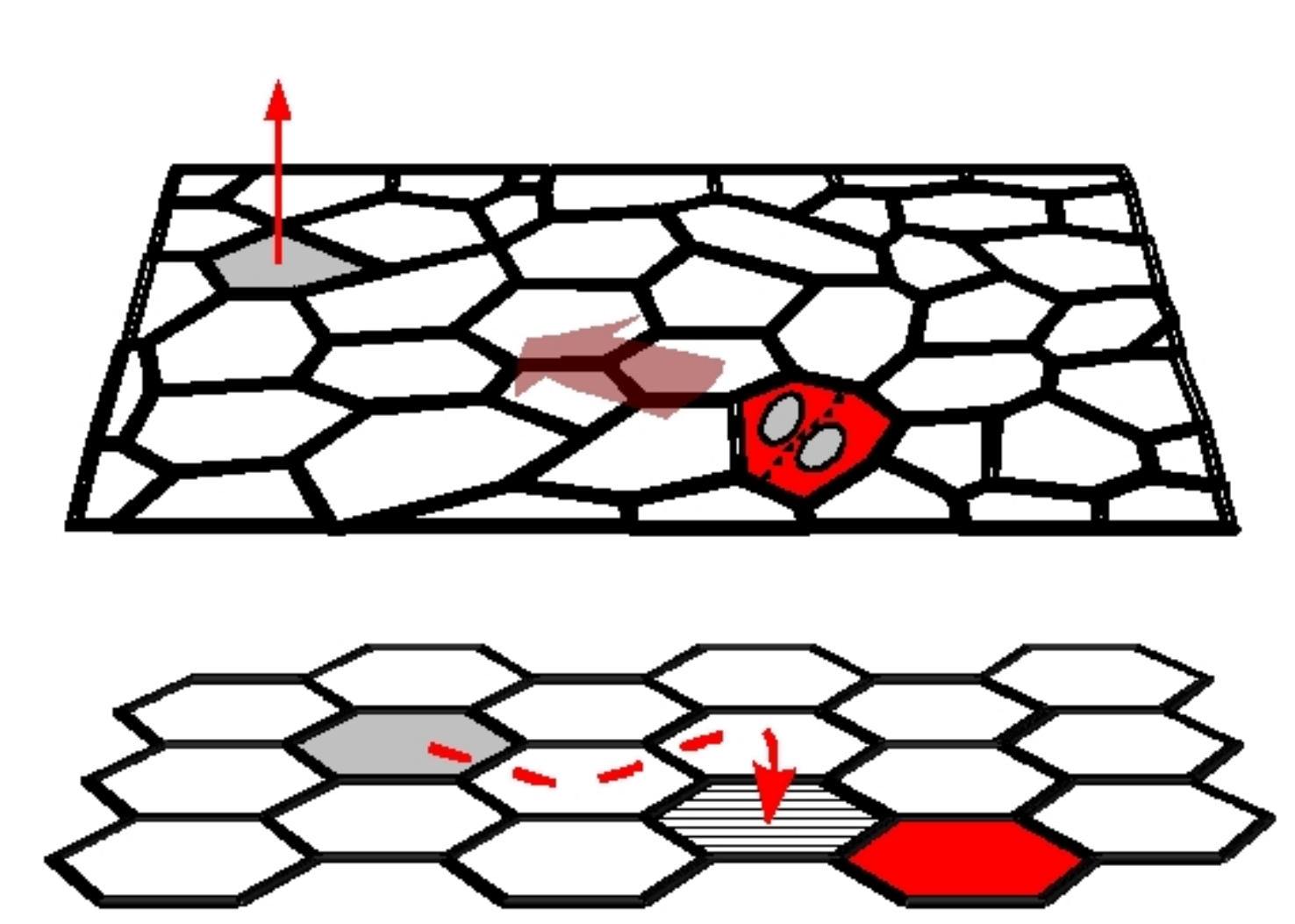}
\caption{\small 
(Color online) 
Schematic motivating the proposed lattice model defined in Eq. (\ref{eqn:2DModel}). 
\emph{Top:} A cartoon of cells within the basal layer, showing the exit
of a cell (light grey) through upward migration into the
suprabasal layers, concurrent with the division of a nearby progenitor cell (red online). To ensure continuity of the basal layer, it is postulated that cells 
rearrange to maintain a uniform cell density (wide arrow).
\emph{Bottom:} In a simple lattice model that captures the essence of
the steady-state dynamics, the exit of a cell from the 
basal layer gives rise to a vacant lattice
site (light grey), which rapidly diffuses by exchanging position 
with adjacent cells (white).
Upon coming into contact with a proliferating cell, the latter may divide
and replace the vacancy with a daughter cell (striped).
%(b) Schematic demonstrating the variation in the effective division rate between regions containing cohesive clusters of progenitor cells (left), and fragmented clusters (right). Empty hexagons indicate progenitor cells, and hashed hexagons indicate B cells exiting through upward migration. With an equal number of type A cells and exiting type B cells, the number of progenitor cells likely to undergo division (shown red in online version) is reduced in the former case compared to the latter. Referring to the discussion in section~\ref{sec:voter}, this leads to the opposite behaviour from process (\ref{eqn:model_Voter}), whereby the cells on the rougher interface would divide more frequently than on the smoother one.
}
\label{fig:latticeModel}
\end{center}
\end{figure}
%

%Therefore, it is salient to start by directly mapping the rules in process (\ref{eqn:model_0d}) onto a phenomenological model in which the areal cell density is implicitly constrained. 
%One may then draw upon further observations of clone and basal layer morphology in order to identify additional rules governing cell behaviour. 
To understand how the experimental observations constrain any proposed theory of spatial behaviour in the basal layer, we shall first address the constraint imposed by the observed uniform cell density. To this end, we postulate that proliferating cells only divide upon the migration of a nearby differentiated cell into the supra-basal layers (see Fig.~\ref{fig:latticeModel}). This requirement is purely phenomenological, as it ensures a uniform density without specifying the mechanism by which it is implemented. Indeed, a range of regulatory pathways can be seen to give rise to the same phenomenology, for example by coupling cell division processes to the local stress~\cite{shraiman:05, shraiman:07} or using short-ranged morphogen gradients as well as feedback by cell-cell communication. 
Once this initial constraint is accounted for in this section, we shall draw upon the further observations of clone and basal layer morphology in order to identify additional rules governing cell behaviour. 

In general, it is a challenge to couple the division and migration of basal layer cells while still allowing for some degree of cell compressibility, whereby a dividing cell may compensate for the exit of a non-adjacent cell through lateral motion (see Fig.~\ref{fig:latticeModel}, top). Several approaches have been used in the past to overcome this problem. In the context of tissue development, one may treat the cell tissue as an elastic medium, whereby the local cell density is coupled to the cell division process through the stress in the surrounding tissue~\cite{shraiman:05}. Such an approach is capable of accounting for a range of realistic properties of two-dimensional cell tissue growth, such as the distribution of cell sizes, as well as cell compressibility. Yet, for the simple problem of steady-state tissue maintenance, involving no net growth, it is unlikely that the complexity of the elastic tissue model is required to explain much of the experimental data. A second approach treats the two-dimensional basal layer as a `foam' of cells (a Voronoi tesselation), for which the steady-state condition may be used to relate between the local cell topology (or more precisely, the number of nearest-neighbours in the basal layer) and the likelihood of division or migration. Applied to the problem of epidermal maintenance, this approach successfully predicts the steady-state topology of epidermal basal-layer cells~\cite{rivier:97}, and it identifies that cells with a larger number of neighbours are more likely undergo division, whereas cells with a small number of neighbours are more likely to migrate in to the supra-basal layers. 
However, it is a challenge to extend this model to allow for two distinct cell populations that are exclusively committed to either division or migration, such as described by process (\ref{eqn:model_0d}). Yet a third  approach draws upon simulations in which cells are modeled as quasi-spherical particles that deform during cell division~\cite{Drasdo:95:05}. However, as this approach draws upon a wide range of (uncontrolled) parameters to describe the cell-cell interactions, it is more complex than required for this case. 

Therefore, in the following we shall suffice with a simpler description of the basal layer, by drawing upon the non-equilibrium lattice models discussed in the recent literature. 
In particular, we shall model the basal layer as a lattice in which each site is occupied by one cell. Cell compressibility is then modeled by a %rapidly diffusing 
population of lattice vacancies, which are created upon cell migration ($B\rightarrow \oslash$) and 
%REVISION
then diffuse rapidly as compared to the cell kinetic rates ($\lambda,\ \Gamma$), before annihilating upon the 
%
%are annihilated upon 
division of an adjacent cell (see Fig.~\ref{fig:latticeModel}). Since this rule-based model is not primarily based on a direct physical representation of individual cells, it may overlook certain physical effects. For example, the migration of type B cells out of the basal layer may be facilitated by mechanical forces exerted by neighbouring cells --- a situation that is hard to represent with a cellular automaton. However, recalling that the stochastic rules embodied by process~(\ref{eqn:2DModel}) have been experimentally verified, it is reasonable to start by considering a similar stochastic process in two dimensions. Later, we shall further justify the use of the lattice model by showing that the lattice geometry does not effect the qualitative behaviour of the system (see section~\ref{sec:voter}(e)).

Although a significant advantage of the lattice description of the basal layer is the ease by which it may be simulated, a range of analytic results are also made accessible by showing that in certain limits, the model reduces to a simple two-component model that belongs to the generalised \emph{voter model} universality class~\cite{Ligget:85, Dornic:01, Hammal:05}. The voter model universality class describes lattice processes that undergo phase separation in two dimensions in the absence of surface tension. In the context of basal layer kinetics, this ``phase separation'' corresponds to the clustering of proliferating cells. Curiously, in the special case $r=1/4$, the basal layer lattice model reduces to the reaction-limited \emph{monomer-monomer} model of surface catalytic reactions, in which the classical voter model dynamics are augmented by infinite-temperature Kawasaki exchange dynamics~\cite{Krapivsky:92}. As well as drawing upon a range of existing results afforded by these models, we shall derive here an exact solution for the two-point spatial correlation function of the monomer-monomer model. These results reveal the continuous transition between voter-like behaviour and diffusive behaviour as the relative rates of symmetric vs. asymmetric cell division (or reactant deposition in surface catalysis) are adjusted.

\subsection{The lattice model}
\label{sec:latticeModel}
As mentioned above, we are interested in constructing a spatial model
that recovers the behaviour of process (\ref{eqn:model_0d}), and maintains a uniform cell density.
To account for the steric repulsion of basal layer cells, 
we will characterise the basal layer as a two-dimensional lattice, 
where each site is host to one of the two cell types, or it remains vacant. 
When the vacancy fraction is very low, then such a lattice description
presents a reasonable approximation of the observed
near-uniform arrangement of basal layer cells during normal adult skin maintenance.
Then, to regulate the cell density, progenitor 
cells (A) are allowed to divide only when neighbouring a site vacancy, 
while the migration of post-mitotic 
(B) cells from the basal layer leads to the creation of vacancies which are free to 
diffuse in the basal layer through the displacement of neighbouring cells (see Fig.~\ref{fig:latticeModel}). 
%To incorporate the existing results of clonal analysis,
%we will suppose that cell division may lead to symmetric or asymmetric cell fate
%in accordance to process (\ref{eqn:model_0d}), 
In summary, denoting a site vacancy with the symbol $\oslash$, the lattice model 
may be written in terms of the non-equilibrium process:
\begin{equation}
\label{eqn:2DModel}
\begin{array}{lcl}
{\rm A}\ \oslash & \stackrel{\lambda'}{\longrightarrow} & 
\left\{\begin{array}{ll}
{\rm A\ A} & {\rm Prob.\ } r \\
{\rm A\ B} & {\rm Prob.\ } \frac{1}{2}-r \\
{\rm B\ A} & {\rm Prob.\ } \frac{1}{2}-r \\
{\rm B\ B} & {\rm Prob.\ } r
\end{array}\right.  \\ 
B & \stackrel{\Gamma}{\longrightarrow} & \oslash \\
{\rm X}\ \oslash & \stackrel{\sigma}{\longrightarrow} & \oslash\ {\rm X}\, ,
\end{array}
\end{equation}
where the site hopping rate $\sigma$ reflects the capacity of vacancies to diffuse within the basal layer, and X is used to denote either a type A or a type B cell. 
To gain some initial insight into the dynamics, and to identify constraints on the parameter space, one may consider the steady-state mean-field cell densities associated with this process. It is straightforward to show that the mean-field equation for the the vacancy fraction $n_\oslash$ is given by
\begin{equation}
\label{eqn:n0}
\partial_t n_\oslash = \sigma\nabla^2 n_\oslash + \Gamma (1-\rho-n_\oslash) - \lambda' \rho n_\oslash\, ,
\end{equation}
where $\rho$ is the (constant) A-cell fraction. From here one may see that the near-uniform cell density in the basal layer ($n_\oslash\ll1$ and uniformly distributed) constrains us to the region of parameter space
$\lambda'\gg \Gamma$, 
%\ \sigma \gg \Gamma$, 
such that any vacancy created through the migration of a type B cell out of the basal layer is rapidly removed %through the division of a 
%REVISION
upon coming into contact with a
proliferating cell. 
%REVISION
%through the process of cell division.
%
In this limit, the numerical value of the parameter $\lambda'$ 
becomes irrelevant, as may be seen by calculating the effective local division rate $\lambda \equiv \lambda' n_\oslash$,
which takes the value $\lambda = \Gamma (1-\rho)/\rho$, independent of $\lambda'$. This relationship between the rate of cell division and migration is identical to that obtained in process (\ref{eqn:model_0d})~\cite{Klein:07}.

%REVISION
Although the uniform vacancy density is a stable fixed point of the mean-field dynamics (Eq. \ref{eqn:n0}), one may worry whether fluctuations about the mean-field solution are capable of compromising the uniform density of the basal layer cell lattice in actual manifestations of process (\ref{eqn:2DModel}). To eliminate this concern, one requires rapid dissipation of density fluctuations independently of the cell kinetics, from which we infer a that the lattice vacancy population must diffuse rapidly compared to the time-scale of cell division and upward migration, viz. $\sigma\gg\Gamma$, as stated earlier. Biologically, this condition corresponds to the assumption that cells are largely incompressible, so that local density fluctuations lead to the rearrangement of cells on a time-scale that is significantly faster than that of cell division (cf. Ref.~\cite{shraiman:05}).
In section~\ref{sec:clone} we shall show that the empirical clone fate data further constrain the parameter space to the region $\lambda' \gg \sigma$, whereby the exit of a type B cell is compensated for by a nearby cell division.

With these definitions, in the parameter space $\lambda' \gg \sigma \gg \Gamma$ one may see that the spatial model introduces \emph{no new relevant parameters} compared to process (\ref{eqn:model_0d}). 
%REVISION
That is, the model behaviour depends only on the (known) zero-dimensional parameters ($r$, $\lambda$, $\Gamma$), with the contribution of the new parameters ($\lambda'$, $\sigma$) entering through the dimensionless ratios $\Gamma/\lambda'$, $\Gamma/\sigma$ and $\sigma/\lambda'$, all of which vanish.
But, as a precondition on its validity, does the model also reproduce the observed `zero-dimensional' basal layer clone size distributions? This is by no means obvious, given the critical (and therefore delicate) nature of process (\ref{eqn:model_0d})~\cite{Klein:07}. 
For example, a high density of progenitor cells in the lattice model may lead to vacancy depletion and jamming, an effect that has no analog in the zero-dimensional system.
%For example, an uncontrolled revision of division rates may lead to the population of proliferating cells becoming extinct. 
Therefore, in section~\ref{sec:experiment} we shall show, using Monte-Carlo simulations, that the proposed lattice model indeed succeeds in reproducing the empirical clonal statistics from Ref.~\cite{Clayton:07}.
With this basic confidence in the validity of the model, we now proceed to analyse its behaviour in more detail.

In general, the cell kinetics in process (\ref{eqn:2DModel}) describe a hard-core non-equilibrium system involving three cell species. Recent progress in the field of non-equilibrium statistical mechanics
has resulted in several possible formalisms with which to study such systems \cite{Odor:04, vanWijland:00, Park:05}. However, for the case at hand, these approaches are unnecessarily complex. 
Rather than analysing the current microscopic model, it is convenient to recast the cell kinetics into a simpler form that describes the same phenomenology. 
%Interestingly, the resulting model is connected to the monomer-monomer model of surface catalysis that has been discussed elsewhere in the literature. 
In particular, for the parameter space of interest ($\lambda'\gg\sigma\gg\Gamma$), it is sufficient to consider a lattice fully occupied by A and B cells, without addressing the population of lattice vacancies. To see this, one may see from Eq.~(\ref{eqn:n0}) that the vacancy dynamics occur on a fast time scale compared to that of cell migration ($1/\Gamma$), and division ($1/\lambda$). Therefore, referring back to the lattice process (\ref{eqn:2DModel}), we may (heuristically) eliminate the vacancy population by replacing the cell division process ($\rm A \ \oslash \rightarrow X\ Y$) with a direct cell-cell `reaction' process ($\rm A\ B \rightarrow X\ Y$), and by replacing the A-cell division rate $\lambda'$ with the effective rate $\lambda(\mathbf{x}) = \lambda' n_\oslash(\mathbf{x}) = \Gamma[1-\rho(\mathbf{x})]/\rho(\mathbf{x})$. Here, the local A-cell fraction $\rho(\mathbf{x})$ refers to the A-cell number density coarse-grained over the nearby lattice neighbourhood. For example, denoting the number of type A cells on a lattice site as $n_\mathbf{x}$, we have $\rho(\mathbf{x}) = \sum_{\mathbf{x}'} n_{\mathbf{x}'} w(|\mathbf{x-x}'|)$ where $w(x)$ is some suitably chosen normalised envelope function.
%fraction by its mean-field value of $n_\oslash = \Gamma(1-\rho)/\rho\lambda'$, and identify the A-cell fraction $\rho$ with its local coarse-grained value $\rho = \rho(\mathbf{x})$, where $\rho(\mathbf{x}) = \int\, d\mathbf{x}' \rho(\mathbf{x}') w(\rho(\mathbf{x-x}')$
%In this study, we shall content ourselves with a heuristic definition of the local A-cell fraction $\rho(\mathbf{x})$).
Within this framework, one may then replace process (\ref{eqn:2DModel}) with the more simplified form
\begin{equation}
\label{eqn:model_Voter}
\begin{array}{lcl}
{\rm A\ B} & \stackrel{\lambda(\mathbf{x})}{\longrightarrow} & 
\left\{\begin{array}{ll}
{\rm A\ A} & {\rm Prob.\ } r \\
{\rm A\ B} & {\rm Prob.\ } \frac{1}{2}-r \\
{\rm B\ A} & {\rm Prob.\ } \frac{1}{2}-r \\
{\rm B\ B} & {\rm Prob.\ } r
\end{array}\right. 
\end{array}\, .
\end{equation}
%where the vacancy-mediated division rate $\lambda'$ has been replaced by an effective average density-dependent rate $\lambda(\mathbf{x})$. 
%\footnote{Recall that $\lambda(\mathbf{x}) = \lambda' n_{\oslash}(\mathbf{x})$, which depends on the local density of holes $0<n_{\oslash}(\mathbf{x})<1$. From the mean-field equations one finds the spatially uniform solution $n_{\oslash}(\mathbf{x}) \simeq \frac{\Gamma (1-\rho)}{\lambda' \rho}$, and the approximation follows.}.
The degree to which this heuristic simplification is indeed justified will be discussed at the end of section~\ref{sec:voter}, together with a quantitative comparison of the behaviour of the exact and simplified models. It is already clear that process (\ref{eqn:model_Voter}) cannot describe the explicit upward migration of post-mitotic cells from the basal layer. However, in the physically relevant limit $\lambda'\gg\sigma$, eliminating the vacancy population has no qualitative effect on the statistics of the progenitor cell compartment, and therefore processes (\ref{eqn:2DModel}) and (\ref{eqn:model_Voter}) are expected to result in the same basal layer phenomenology.

%By simplifying the problem from three to two cell types, 
Interestingly, process (\ref{eqn:model_Voter}) is closely related to the model of monomer-monomer surface catalysis~\cite{Frachebourg:96, Hinrichsen:06}. In particular, when the coarse-grained distribution of type A cells is effectively uniform, such that $\lambda(\mathbf{x})\simeq \rm const.$, then one may identify the symmetric branches of process (\ref{eqn:model_Voter}) with the classical zero-temperature \emph{voter model},
while the asymmetric division channel ${\rm A\ B} \rightarrow {\rm B\ A}$, describes the Kawaski dynamics of an infinite-temperature Ising spin model. Later it will become clear that for the empirical value of $r=0.08$, the significant contribution of the latter will justify the approximation of near-constant $\lambda$.

This analogy provides access to several known results. In Ref.~\cite{Krapivsky:92}, Krapivsky showed that starting from random initial conditions, the classical voter model (i.e. with $r=1/2$) will lead to a lattice of $N$ sites becoming completely saturated with either type A or type B cells after a time $t\sim N\ln N$. Moreover, Frachebourg and Krapivski gave an exact solution for the two-point spatial correlations in this case \cite{Frachebourg:96}, from which they inferred that, 
%REVISION
in the time leading up to saturation ($t\ll N\ln N$), 
the different cell types separate into domains of ever-increasing size, with a typical lengthscale growing as $L\sim \ln t$, and with the density of interfaces $c_{\rm AB}$ between type A and type B cells dropping as $c_{\rm AB} \sim 1/\ln t$. 
%REVISION
As the system approaches saturation ($t\sim N \ln N$), one of the cell types comes to dominate.
The classical voter model is an example of domain growth in the absence of surface tension~\cite{Hinrichsen:06}. Therefore, the boundaries between domains rich in A and B cells are completely unstable, leading to strikingly different and irregular domain morphologies, compared to the smooth phase-separated shapes resulting from surface-tension mediated domain growth.

Qualitatively, the results found for the classical voter model ($r=1/2$) allow us to make several interesting predictions relating to the spatial distribution of A and B cells. In particular, some degree of clustering of proliferating cells is to be expected in adult mice, 
%REVISION
resulting from the growth of domains rich in progenitor cells. Moreover, the ongoing growth of the domain size $L$ suggests that larger clusters are 
%
%with more clustering 
expected in old vs. young epidermis. 
Yet, to make full contact between process (\ref{eqn:2DModel}) and the empirical data, it becomes necessary to calculate the model properties whilst allowing for the relatively low value of $r=0.08$ found in the experimental system.
Therefore, in the following, we shall extend the analysis of Frachebourg and Krapivsky to obtain results valid for arbitrary $r$. Indeed, with $r=1/4$, the following analysis results in the exact solution to the reaction-limited monomer-monomer surface catalysis model. 

\subsection{Exact solution for two-point correlations}
\label{sec:voter}

In the following, we will follow the same approach as taken in Refs.~\cite{Krapivsky:92, Frachebourg:96} for the $r=1/2$ case, but we generalise to allow for arbitrary $r$ and different lattice geometries. For completeness, we include here aspects of the solution that were also described in some detail in Ref.~\cite{Krapivsky:92}, such as the Master equation and the dynamical equations required to define the problem.
We start by identifying type A cells with state $1$ and type B cells with state $0$, so that a lattice with site index $\mathbf{i}$ may be described in terms of the Ising variables $\Phi = \{n_\mathbf{i}\}$, $n_\mathbf{i}\in\{0,1\}$. Referring to Ref.~\cite{Krapivsky:92}, the Master equation for the probability distribution $P(\Phi, t)$ for the system to occupy state $\Phi$ at time $t$ is given by,
\begin{widetext}
\begin{eqnarray}
\label{eqn:master}
\frac{d}{dt} P(\Phi, t) & = & \frac{\lambda}{2} \sum_\mathbf{i,\, e} \Bigg\{ r\left[ U^{(\mathbf{e})}_\mathbf{i}(\hat{F}_\mathbf{i}\Phi) P(\hat{F}_\mathbf{i}\Phi, t) + U^{(\mathbf{e})}_\mathbf{i}(\hat{F}_{\mathbf{i}+\mathbf{e}}\Phi) P(\hat{F}_{\mathbf{i}+\mathbf{e}}\Phi, t) \right] \nonumber \\
        &  &  + (\frac{1}{2}-r) U^{(\mathbf{e})}_\mathbf{i}(\hat{F}_\mathbf{i} \hat{F}_{\mathbf{i}+\mathbf{e}}\Phi) P(\hat{F}_\mathbf{i}\hat{F}_{\mathbf{i}+\mathbf{e}}\Phi, t) - (\frac{1}{2}+r) U^{(\mathbf{e})}_\mathbf{i}(\Phi) P(\Phi, t) \Bigg\}.
        % \nonumber \\
%       &  &  + W_{ij}(F_{ij}F_{ij+1}N) P(F_{ij}F_{ij+1}N, t) - W_{ij}(N) P(N, t) \Big\}.
\end{eqnarray}
\end{widetext}
Here, $\{\mathbf{e}\}$ represent the nearest-neighbour lattice vectors ($|\mathbf{e}|=1$), and
$U^{(\mathbf{e})}_\mathbf{i}(\Phi)\in \{0,1\}$ indicates whether the cells at sites $\mathbf{i}$ and $\mathbf{i}+\mathbf{e}$ are a `reactive' pair, viz.
\begin{eqnarray*}
U^{(\mathbf{e})}_\mathbf{i}(\Phi) &=& n_\mathbf{i}+n_{\mathbf{i}+\mathbf{e}} - 2n_\mathbf{i}n_{\mathbf{i}+\mathbf{e}}\, .
\end{eqnarray*}
The ``spin-flip'' operator is defined by $\hat{F}_\mathbf{i}\Phi = \{ n_\mathbf{j} {\rm\ for\ all\ } \mathbf{j} \ne \mathbf{i};\ 1-n_\mathbf{i} \}$, so that $\hat{F}_\mathbf{i}\Phi$ and $\hat{F}_\mathbf{i+e}\Phi$ correspond to the symmetric division channels and $\hat{F}_\mathbf{i}\hat{F}_{\mathbf{i}+\mathbf{e}}\Phi$ corresponds to an asymmetric division in which the location of the type A and B cells is reversed (viz. $\rm A\ B \rightarrow B\ A$).

From here, recalling that 
$\langle n_\mathbf{i} n_\mathbf{j}\rangle = \sum_\Phi n_\mathbf{i} n_\mathbf{j} P(\Phi,t)$, it is simple to show that the two-site correlation function evolves according to the discretized diffusion equation for non-neighbouring sites,
\begin{equation}
\label{eqn:diffn}
\frac{d}{dt} \langle n_\mathbf{i} n_\mathbf{j} \rangle = \frac{\lambda}{2} (\Delta_\mathbf{i} + \Delta_\mathbf{j}) \langle n_\mathbf{i} n_\mathbf{j} \rangle ,
\end{equation}
where $\Delta_\mathbf{i}$ is the discrete Laplacian operator, defined by
$\Delta_\mathbf{i} n_\mathbf{i} =  \sum_\mathbf{e}(n_{\mathbf{i}+\mathbf{e}} - n_\mathbf{i}).$
However, the diffusion equation is modified for nearest-neighbour correlations, giving
\begin{eqnarray}
\label{eqn:moment2Full}
\frac{d}{dt} \langle n_\mathbf{i} n_{\mathbf{i}+\mathbf{e}} \rangle &=& \frac{\lambda}{2} (\Delta_\mathbf{i} + \Delta_{\mathbf{i}+\mathbf{e}}) \langle n_\mathbf{i} n_{\mathbf{i}+\mathbf{e}} \rangle  \nonumber \\
& + &  (\frac{1}{2}-r)\lambda [ 2\langle n_\mathbf{i} n_{\mathbf{i}+\mathbf{e}} \rangle - \langle n_{\mathbf{i}+\mathbf{e}} \rangle  \nonumber \\
& - & \langle n_\mathbf{i} \rangle] ,
\end{eqnarray}
and the on-site moment is trivially $\langle n_\mathbf{i} n_\mathbf{i} \rangle = \langle n_\mathbf{i} \rangle$.

Making the simplifying assumption that the initial distribution $P(\Phi,0)$ is translationally invariant, then the fraction of type A cells is given by $\langle n_\mathbf{i} \rangle = \rho = \rm const.$, and $\langle n_\mathbf{i} n_\mathbf{j} \rangle$ depends on $(\mathbf{i}-\mathbf{j})$ at all times. Therefore, introducing the correlation function $C_\mathbf{i} = \langle n_\mathbf{j} n_{\mathbf{j}+\mathbf{i}} \rangle$, we can rewrite Eqs.~(\ref{eqn:diffn}) and~(\ref{eqn:moment2Full}) as follows:
\begin{equation}
\label{eqn:moment2TI}
\frac{d}{dt} C_\mathbf{i}  = \lambda \Delta_\mathbf{i} C_\mathbf{i} - \sum_{\mathbf{e}}\delta_{\mathbf{i},\mathbf{e}}(1-2r)\lambda (\rho - C_\mathbf{e} )\, ,
\end{equation}
for $|\mathbf{i}|\ge 1$, subject to the constraint $C_{\mathbf{0}} = \rho = {\rm const}$. That is, the correlation function evolves according to a discrete diffusion equation with sink terms at the nearest-neighbour sites and with a fixed boundary condition at the origin. The linear nature of the problem allows one to seek a solution in terms of the relevant Green's function, e.g. $\hat\Delta_\mathbf{i}^{-1} \equiv G_{(i_x,i_y)}(t) = e^{-4\lambda t} I_{i_x}(2\lambda t) I_{i_y}(2\lambda t)$ for a square lattice $\mathbf{i}=(i_x,i_y)$. For uncorrelated initial conditions, viz.
$ C_\mathbf{i}(t=0) = \rho^2 + \delta_{\mathbf{i},\mathbf{0}}(\rho-\rho^2)\, , $
one may write down a general solution in the form
\begin{eqnarray}
\label{eqn:Cmn}
C_\mathbf{i}(t) &=& \rho^2 + (\rho-\rho^2)G_\mathbf{i}(t) \nonumber \\
&+& \sum_{\mathbf{j}} \int_0^t d\tau J_\mathbf{j}(\tau)G_{\mathbf{i}-\mathbf{j}}(t-\tau)\, ,
\end{eqnarray}
where $J_\mathbf{i}(t)$ is the source distribution required to both maintain $C_{\mathbf{0}}= \rm const$, and also to incorporate the sink terms from Eq.~(\ref{eqn:moment2TI}). Thus, $J_\mathbf{i}(t) = 0$ for $|\mathbf{i}|>1$, $J_{\mathbf{e}} = -(1-2r)\lambda (\rho - C_{\mathbf{e}} )$, and $J_{\mathbf{0}} = z\lambda (\rho - C_{\mathbf{e}} )$, 
%is defined by the condition
%\begin{equation}
%\label{eqn:J00}
%\rho = \rho^2 + (\rho-\rho^2)G_{00}(t) - z(1-2r)\lambda \int_0^t d\tau  (\rho - C_{01}(\tau) )G_{01}(t-\tau) +\int_0^t d\tau  J_{00}(\tau) G_{0\,0}(t-\tau)\,
%\end{equation}
where $z$ is the number of nearest neighbours, and the initial conditions imply that $C_\mathbf{e}$ is the same for all nearest neighbours. For now we will explicity consider the square lattice ($z=4$), however the effect of changing lattice geometry will be discussed near the end of this section. 
%(In anticipation, one might wish to note that a modification of the lattice geometry will modify the magnitude of the source term). 
Making use of Eq.~(\ref{eqn:Cmn}), one may write down a set of self-consistent equations for the source terms, viz.
\begin{eqnarray}
\label{eqn:J_selfC}
J_{\mathbf{0}}(t) &=& 4\lambda \bigg\{ \rho - \rho^2 - \int_0^t \big[ J_\mathbf{0}(t-\tau) G_{(0,1)}(\tau)   \\ 
&+& J_{\mathbf{e}}(t-\tau) \left(G_{(0,0)}(\tau) + G_{(0,2)}(\tau) + 2G_{(1,1)}(\tau) \right) \big] \bigg\}\, \nonumber \\ 
\label{eqn:J01}
J_{\mathbf{e}}(t) &=& -\frac{1-2r}{4}J_{\mathbf{0}}(t)\, ,
\end{eqnarray}
which, upon taking the Laplace transform $j(p) = \mathcal{L}[J(t)]$, $g(p) = \mathcal{L}[G(t)]$, gives the expression for the source term,
%\begin{equation}
%\label{eqn:jExact}
%\left(\begin{array}{cc} 
%1+4\lambda g_{01} & 4\lambda h \\
%-(1-2r)\lambda g_{01} & 1 - (1-2r)\lambda h
%\end{array}\right)
%\left(\begin{array}{c} j_{00} \\ j_{01} \end{array}\right)
%=
%\lambda \frac{\rho-\rho^2}{p} \left(\begin{array}{c} 4 \\ -(1-2r) \end{array}\right)\, ,
%\end{equation}
\begin{eqnarray}
\label{eqn:jExact}
%\begin{array}{l}
j_\mathbf{0}(p) &=& \frac{4\lambda(\rho-\rho^2)}{p(1+4\lambda g_{(0,1)} - (1-2r)\lambda h)},% \nonumber \\
%j_{01}(p) &=& -\frac{(1-2r)}{4}j_{00}(p) \, .
%\end{array}
\end{eqnarray}
where we have defined $h(p) = g_{(0,0)}(p) + g_{(0,2)}(p) + 2g_{(1,1)}(p)$ for the square lattice.

Note that in the classical voter model ($r=1/2$), the sink terms vanish ($j_{\mathbf{e}}=0$), and a single source is located at the origin. The calculation may then proceed exactly as described in Ref.~\cite{Frachebourg:96}. On the other hand, for the infinite-temperature Kawasaki dynamics ($r=0$), the source and sink terms create no net correlation ($j_{\mathbf{e}} = -j_{\mathbf{0}}/4$), and they serve only to maintain the stationary state $C_{\mathbf{0}} = \rho$, $C_{\mathbf{i\ne0}} = \rho^2$.

The discussion so far has been exact. We now turn to the long-time asymptotic solution of the correlation function, for which it is sufficient to consider the behaviour of $j_\mathbf{0}, j_\mathbf{e}$ at small $p$ ($p\ll \lambda$). In the following we shall make use of the following expansions:
$$\lim_{p/\lambda\rightarrow 0}  g_{(0,0)}(p) = \frac{1}{4\pi \lambda} \ln(32\lambda/p) + \mathcal{O}[p\ln(p)]\, , $$
and (for the same limit $p/\lambda\rightarrow 0$)
\begin{eqnarray*}
 g_{(0,0)}(p) - g_{(0,1)}(p) & = &  \frac{1}{4\lambda} + \mathcal{O}[p\ln(p)] \\
 g_{(0,0)}(p) - g_{(1,1)}(p) & = &  \frac{1}{\pi\lambda} + \mathcal{O}[p\ln(p)] \\
 g_{(0,0)}(p) - g_{(0,2)}(p) & = &  \frac{1-2/\pi}{\lambda} + \mathcal{O}[p\ln(p)] \, .
\end{eqnarray*}
With these expansions, the source terms take the long-time asymptotic values
\begin{eqnarray}
\lim_{p/\lambda\rightarrow 0} & j_{\mathbf{0}}(p) = & \frac{4\pi\lambda(\rho-\rho^2)}{p\left[ \pi(1-2r) + 2r\ln(32\lambda/p) \right]} \, .
\end{eqnarray}
%As expected, $j_{00}(p)$ is identical to the long-time asymptotic expression found in Ref.~\cite{Frachebourg:96} when $r=1/2$, and the effect of Kawasaki dynamics are explicity included for arbritrary $r$. 
The corresponding long-time behaviour is therefore
\begin{eqnarray}
\label{eqn:JSoln}
\lim_{t\gg 1/\lambda} J_{\mathbf{0}}(t) = \frac{4\pi \lambda (\rho - \rho^2)}{ \pi\left(1 - 2r\right) + 2r\ln(32\lambda t)} \, ,
\end{eqnarray}
giving the expected $1/\ln t$ decay. Now the effect of Kawasaki dynamics becomes clear:
one may see that reducing $r$ from its maximal value of $1/2$ has the opposing effect of weakening the magnitude of the net source ($J_{\mathbf{0}}-4J_{\mathbf{e}}\propto 2r$) while extending the time over which the source decays, $t_r \sim (t_{r=1/2})^{1/2r}$. In the trivial limit $r=0$, the system becomes stationary as expected, whereas when $r=1/2$ one retrieves the same expression as in Ref.~\cite{Frachebourg:96}.

Taken together, Eqs.~(\ref{eqn:Cmn}), (\ref{eqn:J01}) and (\ref{eqn:JSoln}) give the solution for the long-time asymptotic behaviour of the two-point correlation function. With a mind towards experiment, as well as to compare with the known results for $r=1/2$, we now summarise the features of this solution: \\ \\
\paragraph{The density of interfaces $c_{\rm AB}$ between A and B cells drops asymptotically as $c_{\rm AB}\sim 1/2r\ln t$.} \ \\
The density of interfaces between adjacent A and B cells, $c_{\rm AB}(t) \equiv 2(\rho - C_{\mathbf{e}}(t))$, is an order parameter used to describe the transition from an uncorrelated initial state with $c_{\rm AB} = 2(\rho-\rho^2)$, to the jammed absorbing state $c_{\rm AB}(t\rightarrow \infty) = 0$. %(Note that highly ordered states may result in $c_{\rm AB} > 2(\rho-\rho^2)$). 
For $r=1/2$, it was previously shown that $c_{\rm AB}\sim 1/\ln t$. 

%Referring to Eq.~\ref{eqn:Cmn}, one may see that the density of interfaces is given by
%\begin{eqnarray*}
%\frac{c_{\rm AB}(t)}{2} &=&  \rho-\rho^2 
%- \int_0^t e^{-4\lambda \tau} \big\{J_{00}'(t-\tau) I_0(2\lambda\tau) I_1(2\lambda\tau) \\
%&+& J_{01}(t-\tau)[ I_0^2(2\lambda\tau) + I_0(2\lambda\tau) I_2(2\lambda\tau) \\
%&+& 2I_1^2(2\lambda\tau) ] \big\} \, .
%\end{eqnarray*}
%At long times, this expression asymptotes to
%\begin{eqnarray*}
%\lim_{t\gg 1/\lambda} \frac{c_{\rm AB}(t)}{2( \rho-\rho^2)} &=& \bigg( 1 - \frac{\pi \lambda }{\pi\left(1-2r\right) + 2r\ln(32\lambda t)}  \\
%&\times& \int_0^\infty e^{-\tau/t} e^{-4\lambda \tau}  \Big[4 I_0(2\lambda\tau) I_1(2\lambda\tau) \\
%& -& (1-2r)( I_0^2(2\lambda\tau) + I_0(2\lambda\tau) I_2(2\lambda\tau) \\
%&+& 2I_1^2(2\lambda\tau)\Big] \bigg)\, .
%\end{eqnarray*}
%Identifying the integral with elements of the Laplace transform $g(p=1/t)$, it is then straightforward to obtain the exact result:
%\begin{equation}
%\label{eqn:cAB}
%\lim_{\lambda t \gg 1}  c_{\rm AB}(t) = \frac{2\pi (\rho-\rho^2)}{2r\ln(32\lambda t) + \pi\left(1-2r\right)} + \mathcal{O}[\ln t / t]\, .
%\end{equation}
From its definition, we identify the order parameter to be proportional to the source term, viz. $c_{\rm AB}(t) = J_\mathbf{0}/2\lambda$, so that the long-time asymptotic behaviour is found from Eq.~(\ref{eqn:JSoln}).
This expression reveals the continuous transition between Kawasaki dynamics ($r=0$, $c_{\rm AB} = \rm const$) and voter dynamics ($r=1/2$, $c_{\rm AB}\sim 1/\ln t$). In particular, as $t\rightarrow\infty$, the absorbing-state phase transition occurs at $r=0$.

\
\newline
\paragraph{The spatial correlation function $C_{\mathbf{i}}(t)$ decays as $a(t)-b(t)\ln|\mathbf{i}|$ at short distances, and as a Gaussian at long distances.}\ \\
Away from the origin, where a continuum description suffices, then the correlation function depends only on the distance $x=|\mathbf{i}|$, and all sources appear to be located at the origin, viz. $J_{\mathbf{i}}^{\rm (eff.)} = \delta_{\mathbf{i},\mathbf{0}} (J_\mathbf{0} + 4J_\mathbf{e})$. Replacing the discrete problem with a continuous one simplifies the Green's function, with $\lim_{x\gg1} G_\mathbf{i}(t) \rightarrow G(x,t) = e^{-x^2/4\lambda t}/(4\pi \lambda t)$ for a square lattice.
As a consequence, the long-time asymptotic correlation function may be approximated as $\lim_{t\gg 1/\lambda, x\gg1} C_\mathbf{i}(t) = C(x,t)$, with
\begin{equation}
\label{eqn:C(x,t)}
C(x,t) = \rho^2 - \frac{2r (\rho - \rho^2)}{ \pi\left(1 - 2r\right) + 2r\ln(32\lambda t)} {\rm Ei}\left(-\frac{x^2}{4\lambda t}\right)\, ,
\end{equation}
where ${\rm Ei}(x) = - \int_{-x}^\infty dt\, e^{-t}/t$ is the exponential integral. From the small-argument expansion of this integral, we find
\begin{equation*}
\lim_{t\gg 1/\lambda,\ x^2 \ll \lambda t} \frac{C(x,t) - \rho^2}{\rho-\rho^2} = a(t) - b(t)\ln x\, ,
\end{equation*}
with 
$$a(t) = 1 - \frac{1-2r(\pi-\gamma_e-\ln8)}{2r\ln(32\lambda t) + \pi\left(1 - 2r \right)}\, ,$$
$b(t) = 4r/[2r\ln(32\lambda t) + \pi\left(1- 2r \right)]$, and $\gamma_e$ is the Euler constant.

At long times, $C(1,t) \geq C_{\mathbf{e}}(t)$, so one may infer that the correlation function is concave near the origin. Away from the origin, it is useful to define a correlation length $\xi$ to characterise the short-range correlations, viz.
\begin{equation}
\label{eqn:corrLength}
\xi^{-1} \equiv (\rho-\rho^2) \left. \frac{\partial C}{\partial x}  \right|_{x=1} =b \, ,
\end{equation}
corresponding to the typical size of A-cell rich domains growing as $\xi \sim \ln t$. One may see that when $r>0$, a variation in $r$ merely adjusts the correlation length by a constant, in contrast to its effect on the order parameter $c_{\rm AB}$. For $r=0$, the lattice configuration is random and $C(x,t)=\rho^2$, corresponding to an ``infinite'' correlation length.

\
\newline
\paragraph{The time to saturation $T$ of a lattice of $N$ sites is approximately
$ T \sim N(\ln N + 1/(2r) )/\lambda$.}
\ \\
As mentioned earlier, the classical voter model predicts that any finite-sized system inevitably approaches an absorbing state, in which the lattice is either completely saturated by type A cells or else they have become extinct. Fortunately, the time-scale $T$ in which the absorbing state is reached is $T\sim (N\ln N)/\lambda$ for $r=1/2$, which, for a biological system with $1/\lambda\sim 1$ week and $N\gg1$, far exceeds the lifetime of a mammalian organism. 

For arbitrary $r$, the time-scale $T$ may be estimated by repeating the calculation in Ref.~\cite{Krapivsky:92}: The saturation condition is $\sum_{\mathbf{i}}C_\mathbf{i}(T) = N\rho$. Replacing the summation by integration, $\sum_\mathbf{i} C_\mathbf{i} \rightarrow \int_0^{\infty} x C(x,t) dx$, we arrive at the asymptotic relation
$$\lim_{\lambda t \gg 1} \sum_\mathbf{i} \frac{C_\mathbf{i}-\rho^2}{\rho-\rho^2} \simeq \frac{4r\lambda T}{\pi(1-2r)+2r\ln(32\lambda T)}\, ,$$
and the result for $T$ follows. 
Interestingly, this result predicts that in even a reasonably large system and finite $r$, when $N\gg e^{\pi/2r}$, then the time to saturation is insensitive to the value of $r$. Indeed, for such systems the cross-over to non-voter-like behaviour only occurs at very small values of $r\sim 1/\ln N$.

\ \newline
\paragraph{The product of the correlation length and interface density gives a time-invariant characteristic of voter-like coarsening.}
\ \\
It has been previously noted that in voter-like coarsening, the characteristic length-scale of domains is inversely proportional to the interface density $c_{\rm AB}$ \cite{Dornic:01}. An important implication of this observation is that at asymptotically long times, one may identify a time-invariant characteristic of the correlations, which we define as $\Omega \equiv (\rho-\rho^2)/(c_{\rm AB}\xi) = 2r/\pi$. 

To characterise the $\Omega$-constant, one may identify its definition as the ratio between the domain size $\xi$ and domain circumference, as calculated from the number of A-B interfaces associated with each domain $c_{\rm AB}\xi^2$. Thus, the domain perimeter has a trivial fractal dimension of one, with $\Omega$ indicating the perimeter roughness, or curvature. 
Values of $\Omega\simeq1$ may be associated with cohesive and smooth domains (i.e. with a large area-to-interface ratio), whereas systems with $\Omega\rightarrow0$ have highly fragmented, or rough, domains. Not surprisingly, smaller values of $r$ lead to rougher domains as a result of the Kawasaki dynamics. Yet, one may see that even at the maximum value of $r=1/2$, the voter model predicts rough domains ($\Omega<1/2$)
% compared to those of the cohesive circular structures, which may be seen in low-temperature surface-tension driven coarsening 
--- an observation readily seen in simulations, see e.g. Ref.~\cite{Hinrichsen:06}. 

%It is interesting to compare the $\Omega$-constant with the value of $\Omega_0 = \pi/4$ associated with cohesive circular domains (or  $\Omega_{||} = 1$ associated with stripe-like domains): one may see that even at the maximum value of $r=1/2$, the voter model predicts lower saturation (and rougher domains) --- an observation readily seen in simulations, see e.g. Ref.~\cite{Hinrichsen:06}. This is a zero-temperature effect intrinsic to the voter model.

With an eye to the empirical analysis in the next section, let us note that the $\Omega$-constant allows us to characterise \emph{static} observations of type A cell correlations, and thus provides a valuable test of whether a given data set is consistent with the long-time behaviour of process~(\ref{eqn:model_Voter}).
\
\newline
\paragraph{The lattice geometry influences only the timescale of clustering, with higher coordination number corresponding to a faster timescale.}
\ \\
For lattice geometries with different coordination number $z\neq4$ (and one site per unit cell), the calculation proceeds as for a square lattice, but now three modifications must be made: First, as mentioned above, the central source term is $J_\mathbf{0} = z\lambda(\rho - C_\mathbf{e})$, and there are now $z$ sink terms $J_\mathbf{e}$ at nearest neighbouring sites. Second, the appropriate Green's function now has the (non-separable) form
$$G_\mathbf{i}(t) = \frac{1}{4\pi^2}\int d^2\mathbf{q} \exp\left[ i\mathbf{q\cdot i}-\lambda t \left(z-\sum_\mathbf{e} e^{i \mathbf{q\cdot e}}\right)\right]\, ,$$
from which we obtain the general form of the small-$p$ (long-time) expansion of the Laplace transform $\lim_{p/\lambda \rightarrow 0} g_{\mathbf{0}}(p) \sim \ln (\lambda/p) + d_\mathbf{0}/2\lambda$ and $g_{\mathbf{i}}(p) = g_{\mathbf{0}}(p) - d_\mathbf{i}/2\lambda$, where $d_\mathbf{i}$ are numerical constants, e.g. $d_\mathbf{e}=2/z$. Third, the sum over sink terms in Eq.~(\ref{eqn:J_selfC}) is revised to reflect the lattice geometry. With these three modifications, one finds that the order parameter $c_{\rm AB}$ takes the general form $c_{\rm AB} = 2\pi(\rho -\rho^2)/[ \alpha(1-2r) + 2r\ln(\beta \lambda t) ]$, where $\alpha$ and $\beta$ are geometry-dependent constants. One is led to conclude that the lattice geometry only serves to rescale the time variable by some constant, $t\rightarrow \beta e^{\alpha(1-2r)} t$. It is interesting to contrast this with the more significant effect of modifying the branching ratio $r$, which instead rescales $\ln t$.

With regards to the experimental system, the apparent insensitivity of the results to details of the lattice geometry reinforces the validity of a lattice-based description of the basal layer. Namely, while it is clear that the basal-layer is not a periodic lattice of uniform cell size and coordination number, it may nonetheless be modeled as such.

Finally, because the average coordination number of $z=6$ is expected for the biological system, we have calculated the order parameter exactly for a hexagonal lattice, giving $\alpha^{\rm (hex)} \simeq 0.98 \pi$, and $\beta^{\rm (hex)} = 51$.

\ \newline
\begin{figure}
\begin{center}
\includegraphics[width=3.3in]{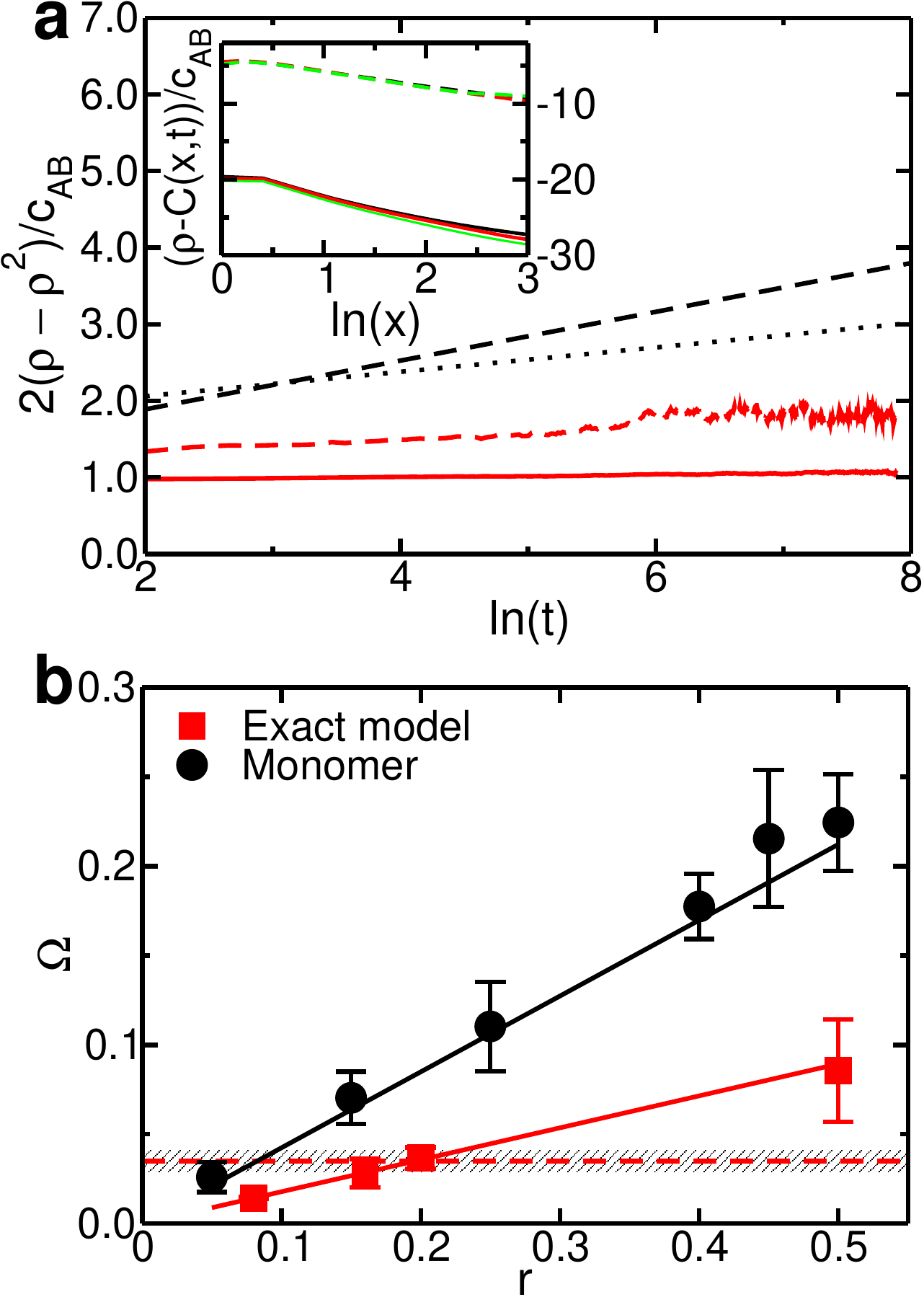}
\caption{\small 
Comparison of the exact basal layer lattice model (\ref{eqn:2DModel}) (red curves) with the simplified model (\ref{eqn:model_Voter}) (black). (a) The order parameter $1/c_{\rm AB}$ is plotted against $\ln t$. The curves correspond to the simplified model with $r=1/2$ (dashed black) and $r=1/4$ (dotted), and to the exact model with $r=1/2$ (dashed red) and $r=0.08$ (solid curve). For the simplified model, the analytical expression for $c_{\rm AB}$ was used. For the exact model, the results were obtained from numerical simulation on hexagonal lattices of size $N=1024\times1024$, with $\rho=0.22$ and setting $\lambda=1$. \emph{Inset}: The correlation function plotted against logarithm of the distance $\ln x$, evaluated from numerical simulation and Eq.~(\ref{eqn:corrFn}). The time-invariant ratio $(\rho-C(x,t))/c_{\rm AB}$ is plotted at $\lambda t=50,100,1000$ for the classical voter model (dashed), and at $\lambda t = 350,3500,7000$ for the exact model with $r=1/2$ (solid curves). Consistent with the expected behaviour (see Eq.~(\ref{eqn:C(x,t)})), the curves overlap and are convex near $x=1$, becoming linear at $x>1$.
(b) The long-time asymptotic roughness constant $\Omega$ plotted against $r$ for both models. 
Data points correspond to the values calculated using Eq.~\ref{eqn:corrFn} from numerical simulations such as shown in Fig.~\ref{fig:CAExample}, using the algorithm described in section~\ref{sec:clone}. Error bars result from fluctuations due to finite-size effects, as evaluated by considering the random variation in $\Omega$ over the time of the simulation. The black curve gives the theoretical value of $\Omega$ for the simplified (monomer) model on a hexagonal lattice, while the red curve gives the best-fit to the numerical results for the exact model.
}
\label{fig:voter}
\end{center}
\end{figure}
This completes our theoretical discussion of process~(\ref{eqn:model_Voter}), and of the monomer-monomer surface catalysis model.
It now remains to be seen whether the exact model of cell division~(\ref{eqn:2DModel}) is indeed described by the properties of the approximate process (\ref{eqn:model_Voter}) (with uniform $\lambda$). Let us recall that the latter is a reasonable approximation assuming that the density of type A cells is approximately uniform. Thus, although the approximation must fail on the time-scale $\lambda T\sim N\ln N$ associated with the jamming transition, the logarithmically slow growth of correlations suggests that at shorter times ($t\ll T$) the degree of progenitor cell clustering should be sufficiently low as to make the analysis self-consistent. 

\begin{figure}
\begin{center}
\includegraphics[width=3.3in]{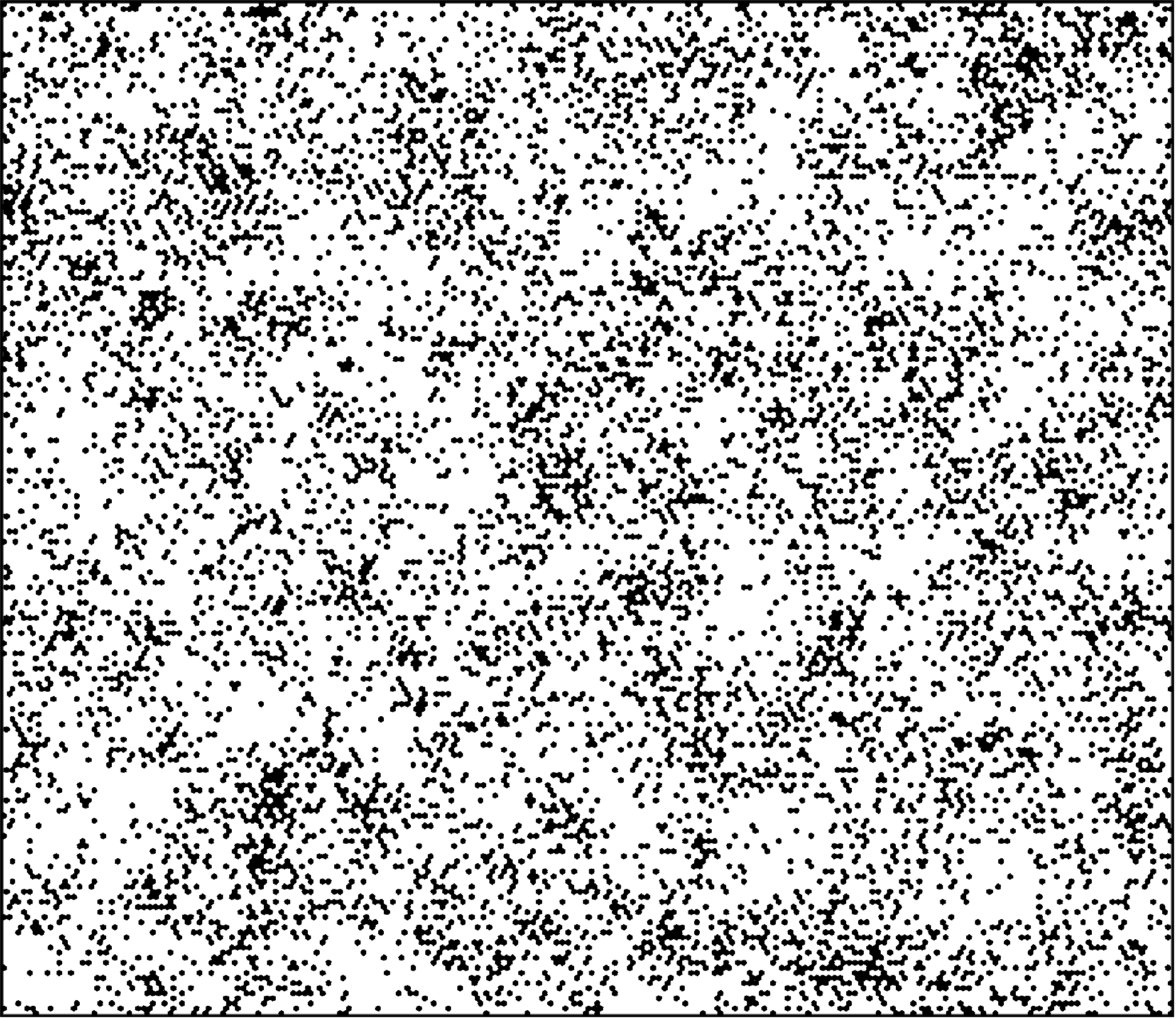}
\caption{\small 
Cellular automaton simulation of process (\ref{eqn:2DModel}), showing the distribution of progenitor cells (black hexagons) on a lattice of $N=200\times200$, and using the experimental branching ratio $r=0.08$ and progenitor cell fraction $\rho=0.22$. The frame shown corresponds to an evolution time of $t=30/\lambda$, where $t=0$ corresponds to random initial conditions. White areas are fully occupied by post-mitotic cells.
}
\label{fig:CAExample}
\end{center}
\end{figure}

To test whether the model indeed satisfies the expected behaviour, in Fig.~\ref{fig:voter} we compare the spatial correlation of type A cells, as given by the properties derived above, with the results of process (\ref{eqn:2DModel}) as obtained by numerical simulation using a Gillespie-like algorithm (described below in section~\ref{sec:clone}). An example of the cellular automata simulations used for the comparison is shown in Fig.~\ref{fig:CAExample}, where the distribution of type A cells (black) is shown on a hexagonal lattice. Qualitatively, one may see from this figure that process (\ref{eqn:2DModel}) does indeed give rise to some clustering. By extracting the correlation function from such figures, we obtained the quantitative comparison shown in Fig.~\ref{fig:voter}.
In Fig.~\ref{fig:voter}a, we compare the evolution of the inverse order parameter $1/c_{\rm AB}$ against $\ln t$ for the two models. One may see from the linear behaviour of the exact model (red online), that the order parameter indeed shows the expected $1/\ln t$ decay. Equally, we may confirm that the radial correlation function $C(x,t)$ has the functional form $a-b\ln x$, with parameters $(1-a), \ b$ both proportiaonal to $c_{\rm AB}$, as demonstrated by plotting $(C(x,t)-\rho)/c_{\rm AB}$ in Fig.~\ref{fig:voter}a (inset). Finally, we may confirm that both models have the same long-time $r$-dependence by plotting $\Omega$ against $r$ in Fig.~\ref{fig:voter}b. Here, one may see that indeed $\Omega$ depends linearly on $r$. 

\begin{figure}
\begin{center}
\includegraphics[width=3.3in]{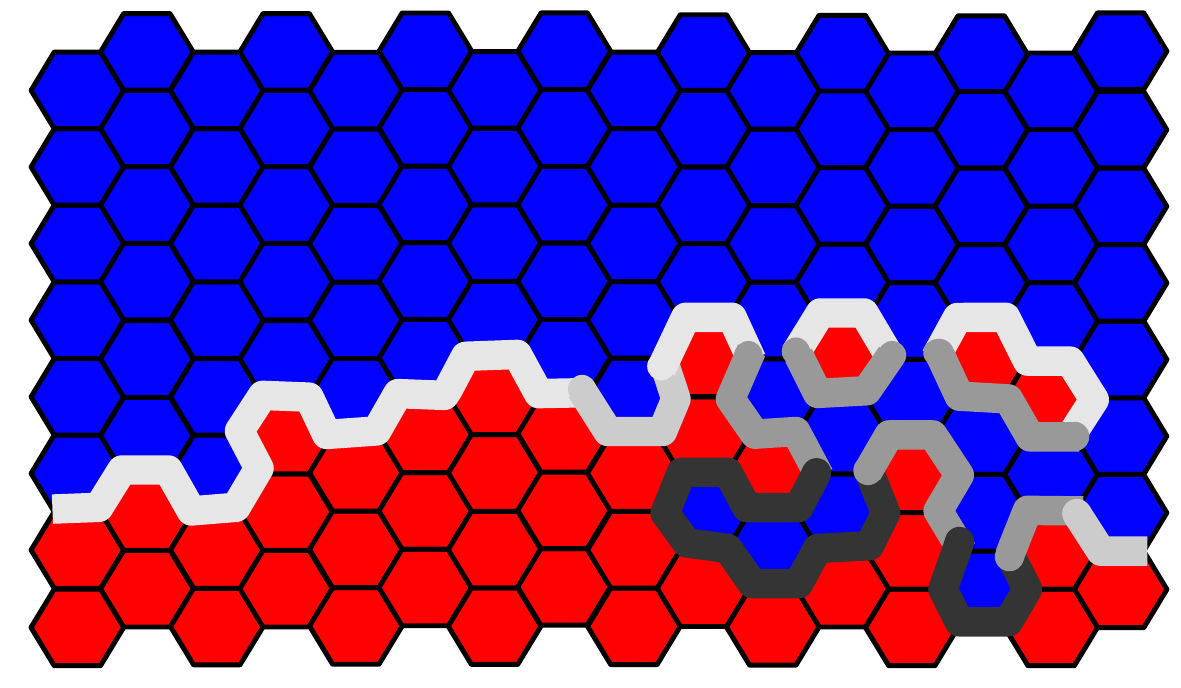}
\caption{\small (Color online)
Schematic demonstrating the variation in the effective division rate in process (\ref{eqn:2DModel}), accounting for the quantitative variation in behaviour between processes (\ref{eqn:2DModel}) and (\ref{eqn:model_Voter}). The reactive interface between progenitor cells (light grey, red online) and post-mitotic cells (dark grey, blue online) is indicated by the thick grey line, with a faster division rate shown in light grey and slower division rate shown in dark grey. On the left, a smooth interface results in faster cell division, as few progenitors on the boundary must compensate for the migration of many post-mitotic cells in the bulk. On the right, the ``rough'' interface results in a variation between fast and slow division rates, due to restricted access of vacancies into the rough interfacial regions.  Referring to the discussion at the end of section~\ref{sec:voter}, this leads to increased stability of rough compared to smooth interfaces.
}
\label{fig:vacancyEffect}
\end{center}
\end{figure}

Yet, although Fig.~\ref{fig:voter} reveals that processes (\ref{eqn:2DModel}) and (\ref{eqn:model_Voter}) result in the same functional dependence of the correlation function on $r$ and $t$, it is striking that the exact model (\ref{eqn:2DModel}) results in a slower growth in correlations compared to the simplified model, characterised both by a slower decay of $c_{\rm AB}$ seen in Fig.~\ref{fig:voter}a, and by rougher domains (or lower $\Omega$) seen in Fig.~\ref{fig:voter}b. How might one explain this change? Referring to the discussion in section~\ref{sec:latticeModel}, let us recall that the simplified model is connected to the exact model by relating the local division rate to the mean-field vacancy density ($\lambda(\mathbf{x}) = \lambda' n_\oslash(\mathbf{x})$), with the two models becoming equivalent when $n_\oslash(\mathbf{x})= \rm const$ for an arbitrarily small degree of coarse-graining. This condition is satisfied in the limit $\sigma \gg \lambda'$ with the process $\rm A\ \oslash \rightarrow \oslash\ A$ removed. However, for the physically meaningful case $\sigma \ll \lambda'$ one can no longer treat the vacancy density as uniform, leading to quantitative (but not qualitative) differences in the behaviour of the two models. As shown schematically in Fig.~\ref{fig:vacancyEffect}, the vacancy population has the effect of accelerating cell division on the edge of smooth progenitor cell clusters, as a result of the higher concentration of nearby post-mitotic cells migrating into the super-basal layers. Conversely, cell division on rough cluster edges is slowed down. In total, rough A-B interfaces remain stable over a longer period of time, leading to the observed differences between the two models.

\section{Empirical analysis}
\label{sec:experiment}
We are now in a position to turn to the experimental analysis of the IFE. In order to identify the underlying rules of cell division and differentiation, one can envision two types of experiment: 
\begin{itemize}
\item First, one may track the fate of \emph{individual cells} and their progeny, and then look for a cell kinetic description compatible with their observed behaviour. Such an approach was used with considerable success in the definition of the zero-dimensional process (\ref{eqn:model_0d}) through clonal analysis, and we shall extend it to the analysis of the spatial process in section~\ref{sec:clone}. %However, the introduction of spatial degrees of freedom greatly complicates the analysis of cell kinetics. 

\item Second, in section~\ref{sec:corr} we shall consider the statistics of the \emph{entire population} of basal layer cells. In particular, by immunostaining basal layer cells for markers of cell proliferation, it is possible to analyse the spatial distribution of all progenitor cells within the layer. Then, referring to the theoretical discussion in section~\ref{sec:model}, one may look for a signature of the underlying cell kinetics in the spatial correlation of proliferating cells. 
\end{itemize}

Note that the two types of experiment give access to independent aspects of cell behaviour: The former probes the temporal evolution of cell lineages, whereas the latter reveals the static basal layer morphology. As such, the experiments provide a significant degree of mutual verification of any proposed theory of cell behaviour. Yet, even the best of such experiments leave room for some ambiguity: For example, it is far from clear what importance should be assigned to the embryonic development of the IFE in predetermining the spatial distribution of cells.
%and to what extent it is valid to think about the initial conditions being random. 
Moreover, one may in principle conceive of regulatory pathways that leave no signature on the spatial distribution of cells, or which may not be distinguished from an independent stochastic process. Thus, in the following we will look for the simplest possible model of cell behaviour that succeeds in capturing the known biological constraints.
%, and which has the capacity to predict cell behaviour. 
%As such, our starting point lies in the proposed lattice model (\ref{eqn:2DModel}).

\subsection{Clonal analysis}
\label{sec:clone}
\begin{figure}
\begin{center}
\includegraphics[width=3.3in]{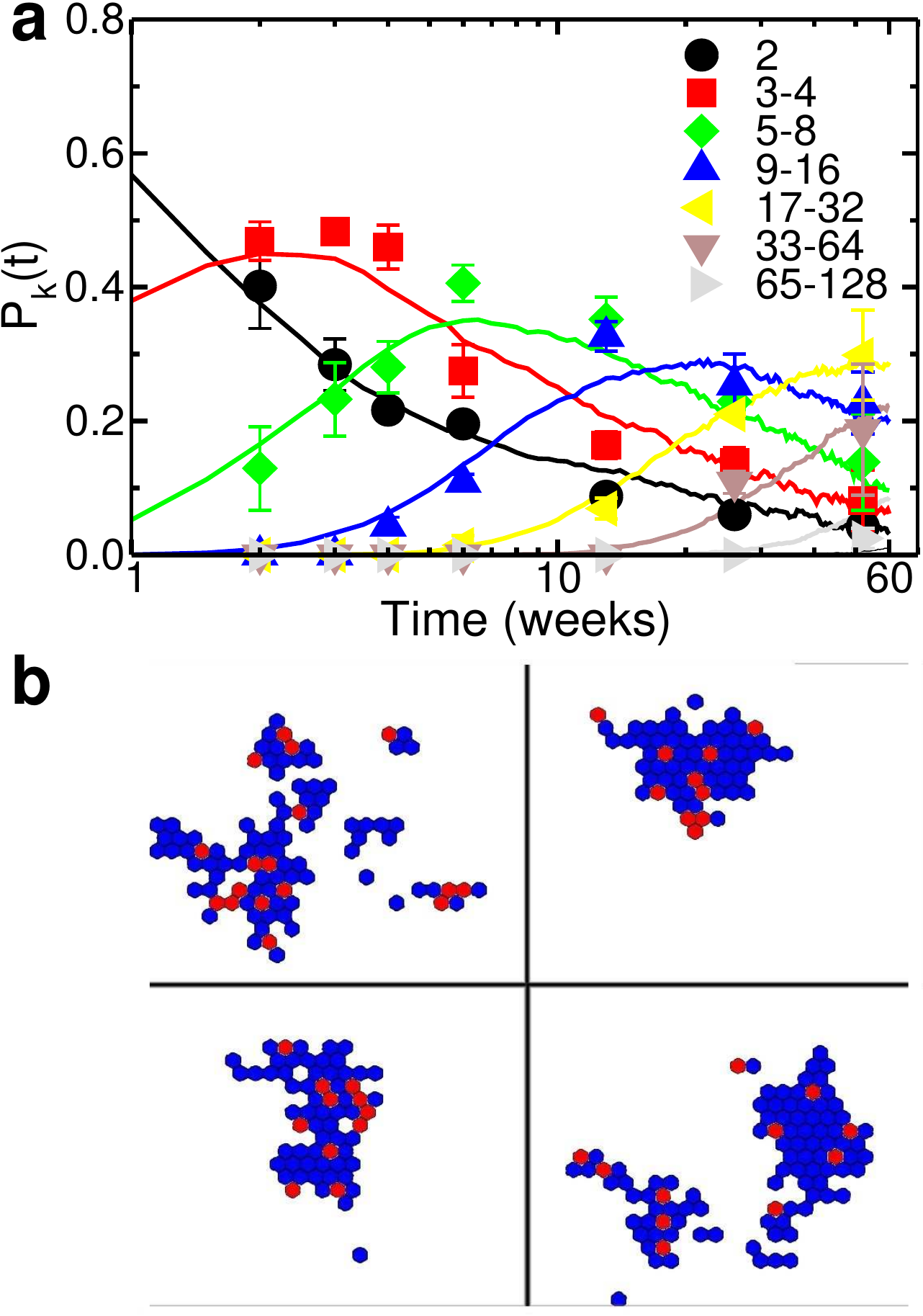}
\caption{\small (Color online)
(a) Comparison of the empirical clone size distribution (data points) to predictions of process (\ref{eqn:2DModel}) (solid curves), as obtained from Monte-Carlo simulations of $10^4$ labelled clones. 
The distributions are plotted in terms of the probability $\mathcal{P}_k(t)$ for a clone to have between $2^{k-1}+1$ and $2^k$ basal layer cells at time $t$ after labelling, normalised to include only clones with 2 or more cells in the basal layer ($k\ge 1$). The empirical data is reproduced from Ref.~\cite{Clayton:07}.
%Simulations were conducted using a hexagonal lattice of $60\times60$ cells, initially seeded at random with a fraction of $\rho=0.22$ sites occupied by type A cells.
%
(b) Examples of the basal layer structure of several large late-stage clones evolving according to process (\ref{eqn:2DModel}), starting from a uniform random distribution of unlabelled cells, with one single type-A cell labelled at $t=0$. Light and dark grey hexagons (red and blue online) indicate sites occupied by type A and B cells, respectively. White areas are populated by unlabelled cells. Each frame corresponds to the progeny of one initially labelled cell. %The simulations used the same parameter set as described in the caption of Fig.~\ref{fig:CAavgs}, but using a lattice of size $N=60\times60$ sites.
}
\label{fig:clone}
\end{center}
\end{figure}
To begin our analysis of the empirical data, we start by considering the fate of individual labelled cells and their progeny, hereafter referred to as clonal fate data.
In an extensive experiment reported elsewhere~\cite{Clayton:07}, the low-frequency labelling of approximately 1 in 600 basal-layer epidermal cells at a defined time was achieved by a drug-inducible genetic event, which resulted in expression of the enhanced Yellow Fluorescent Protein (EYFP) gene in a cohort of mice. At intervals, the EYFP label was detected by confocal microscopy, which enables 3D imaging of entire sheets of epidermis. 
By analysing samples of mouse epidermis at different time points, it was possible to 
analyse the fate of labelled basal layer clones at single cell resolution \emph{in vivo} 
for times up to one year post-labelling in the epidermis (see, for example, Fig.~\ref{fig:skinCrossSection}b) \cite{Braun:03, Clayton:07}. 

%As one may see in Fig.~\ref{fig:clone}, t
The distribution of cells within a clone constitutes a valuable data set, which one may analyse for signatures of the underlying rules of cell division. By scoring the total number of labelled cells $n$ in each clone at progressive time points, it was possible to access the evolution of the full clone size distribution, as given by the probability $P_{n>0}(t)$ for finding a clone with $n>1$ basal layer cells at a time $t$ post-labelling, shown in Fig.~\ref{fig:clone}(a). The properties of the size distribution were used to infer the laws summarised in process (\ref{eqn:model_0d})~\cite{Clayton:07, Klein:07}. In particular, it was shown that in the long-time limit ($t\gg1/r\lambda$), the basal layer clone size distribution conforms to the scaling form
$$\lim_{r\lambda t\gg1} P_{n>0}(t) = \frac{\rho}{r\lambda t}\exp\left(\frac{\rho\, n}{r\lambda t} \right)\, .$$
As discussed in section~\ref{sec:model}, 
%to proceed from a `zero-dimensional' model of cell division to a model that describes the spatial behaviour of cells, 
we must first establish, for any proposed model of spatial behaviour, that the zero-dimensional clone size distributions are faithfully reproduced. 

To this end, we conducted %Monte-Carlo 
multiple 
simulations of process~(\ref{eqn:2DModel}) as an asynchronous cellular automata evolving on a hexagonal lattice of $N=60\times60$ sites over a period 
%REVISION
corresponding to $T=60$ weeks in the experimental system (recall that $\lambda=1.1$/week~\cite{Clayton:07}).
At $t=0$, the lattice was fully occupied by randomly-placed type A and type B cells. A single, randomly chosen, type A-cell was assigned a hereditary `label' at the start of each simulation. In effect, each such simulation mimics the evolution of one labelled clone, 
%REVISION
so that repeated simulations may be used to sample the full clonal statistics, viz. Monte-Carlo sampling. In particular, 
%
%Thus, 
by tracking the number of labelled cells as a function of time over $10^4$ such clone simulations, we could compare the clonal statistics predicted by process (\ref{eqn:2DModel}) with those expected from the zero-dimensional process (\ref{eqn:model_0d}). In keeping with the empirical fit from the previous section, we used a B-cell migration rate of $\Gamma=0.31$/week, and an initial A-cell fraction of $\rho=0.22$. The `fast' rates of cell division and hole diffusion were set to be $\lambda'=10^4$/week and $\sigma=200$/week, although the precise values are unimportant provided that $\lambda'\gg\sigma\gg \Gamma$, as discussed in section~\ref{sec:model}.
The results are plotted in Fig.~\ref{fig:clone}(a), where we compare the clone size distribution $P_{n>0}(t)$ to the empirical data. One may see that the fit is remarkably good, and, within the current empirical resolution is indistinguishable from the predictions of zero-dimensional model (cf. Ref.~\cite{Klein:07}). 
In summary, the fit provides a first validation of process (\ref{eqn:2DModel}) as a viable model of spatial behaviour in the basal layer.

We may now revisit the clone fate data with a view to study spatial structure. Unfortunately, although the clone size data was stored for all clones, images showing their spatial structure were retained only for a small number of clone samples. Therefore we are not in a position to conduct a comprehensive \emph{quantitative} analysis of clone shape evolution.
%Nevertheless, a sufficient amount of clone fate data is available to allow qualitative analysis. 
Nevertheless, a striking qualitative feature of clones is that they remain largely cohesive, as shown in the example in Fig.~\ref{fig:skinCrossSection}b. (Indeed, without this property the very enterprise of clonal analysis would have proved difficult). Therefore, to challenge the validity of the proposed lattice model of cell division, we test, using additional Monte-Carlo simulations, the ability of the model to produce cohesive clones over the one-year time period of the experiment. 
Representative results are shown in Fig.~\ref{fig:clone} for large clones at 60 weeks post-labelling. When $\lambda'\gg\sigma\gg\Gamma$, clones were seen to remain largely cohesive throughout the period of the simulation.
We may therefore conclude that, at least with respect to the existing clonal fate data, process~(\ref{eqn:2DModel}) presents a reasonable phenomenological description of the spatial behaviour of basal layer cells. In particular, the cohesive nature of clones may be completely explained in terms of an independent stochastic process, with no evidence for further forms of regulation.

\begin{figure}[b]
\begin{center}
\includegraphics[width=3.3in]{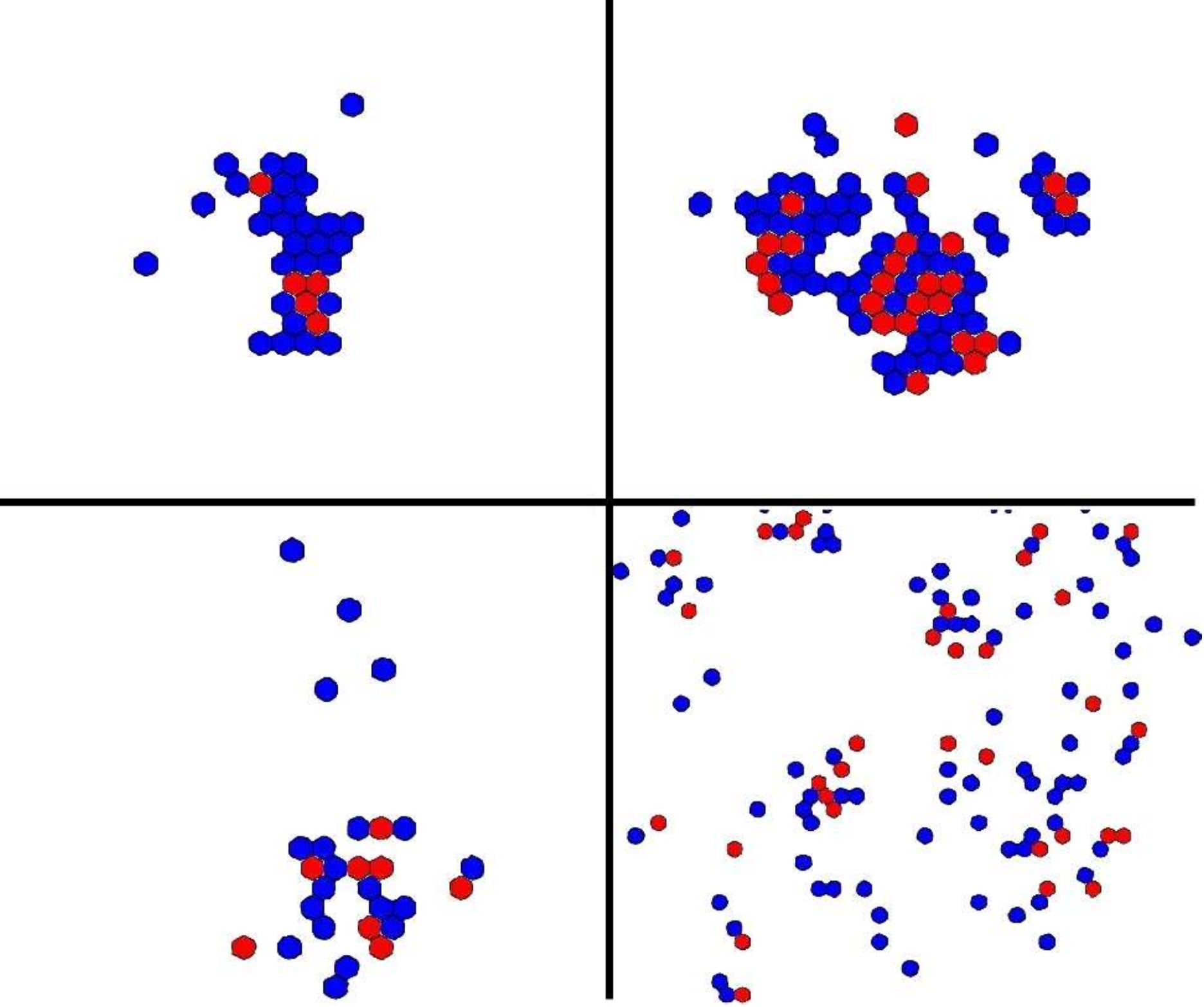}
\caption{\small (Color online)
The loss of cohesiveness with increasing the relative hopping rate $\sigma/\lambda'$ is shown through examples of late-stage clone simulations. Light and dark grey hexagons (red and blue online) indicate sites occupied by type A and B cells, respectively, from the same clone. White areas are populated by unlabelled cells. The simulations used the same parameter set as described above for the clones in Fig.~\ref{fig:clone}, but using the parameter values: $\sigma =20$/week, $\sigma =200$/week with $\lambda'=10,000$/week for the top-left and top-right clones respectively; $\sigma =200$/week, $\sigma =2000$/week with $\lambda'=200$/week for the bottom-left and bottom-right clones. For the latter, which is clearly unphysical, the scale has been reduced twofold to demonstrate the wide dispersion of labelled cells. The examples demonstrate that clones are cohesive when $\sigma/\lambda'\ll 1$, but dispersive otherwise.
}
\label{fig:cloneDispersive}
\end{center}
\end{figure}
Yet, to what extent is the observation of cohesiveness sensitive to modifications of the basal layer lattice model? To test the degree to which cohesiveness constrains the model, we 
allowed for a degree of cell mobility within the basal layer, by introducing the additional exchange process $\rm A B \rightarrow B A$ with rate $\gamma\sim \Gamma$. The exchange process is a reasonable candidate for cell behaviour, motivated by the observation that keratinocytes in culture are highly motile, which leads one to postulate whether cells \emph{in vivo} are capable of independent lateral migration in the basal layer. We found that for any non-small value of $\gamma$, (i.e. $\gamma \gtrsim \Gamma$), the progeny of labelled clones rapidly dispersed, as shown in Fig.~\ref{fig:cloneDispersive}. We are led to conclude that \emph{epidermal cells in vivo move only in response to a local density gradient} resulting from cell division and migration.
Moreover, a second investigation in which the hard-core mobility $\sigma$ was made larger than $\lambda'$ (and keeping $\gamma=0$) again led to a loss of clone cohesiveness. We are therefore led to conclude that, within the framework of the non-interacting lattice model, \emph{progenitor cells undergo division rather than lateral migration as a response to a local drop in density}.

In conclusion, the observation of clone cohesiveness imposes a severe constraint on cell behaviour, which allows us to rule out several variations of the basic lattice model. Yet, at least qualitatively, there is no clear evidence either \emph{for} or \emph{against} additional regulation of cell division in the clone shape data. So to what extent is the lattice model truly %robust? 
%REVISION
capable of shedding new light on the mechanism of cell fate regulation? 
To do better, we must look for quantitative data for comparison. Fortunately, such data can be found in the readily-available spatial distribution of progenitor cells.

%%%%%%%%%%%%%%%%%%%%%
%%%%

\subsection{Correlation analysis}
\label{sec:corr}
\begin{figure}
\begin{center}
\includegraphics[width=3.3in]{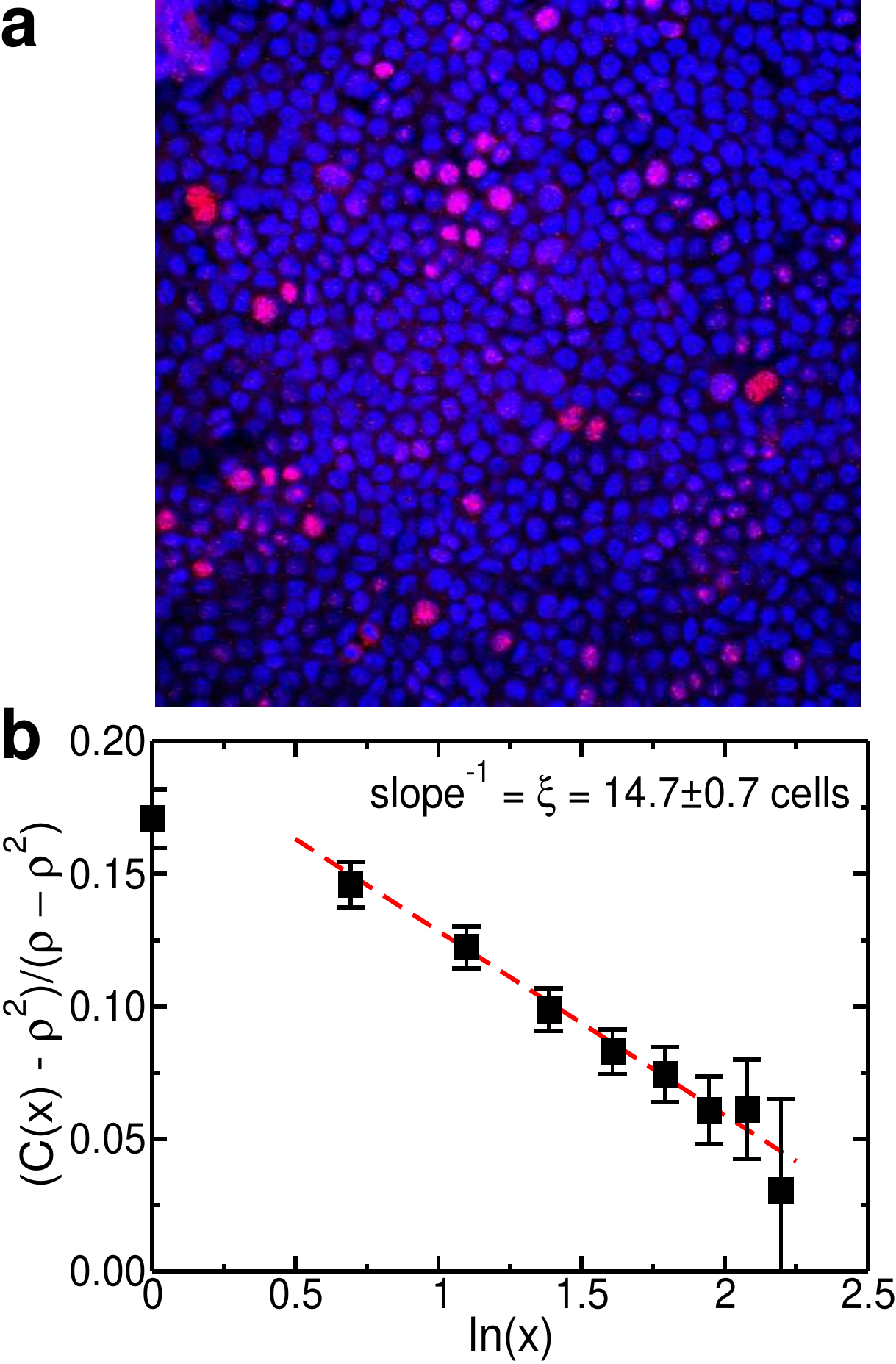}
\caption{\small (Color online)
(a) Confocal micrograph of wholemounted mouse tail skin IFE, showing the two-dimensional basal layer immunostained for the nuclear marker DAPI (dark grey, or blue online) and the proliferation marker Ki67 (light grey, or red online). The area surrounding the stained nuclei is occupied by unstained cell cytoplasm (black).
(b) The empirical radial correlation function $C(x)$, as defined in Eq.~(\ref{eqn:corrFnExp}). Data points show results obtained by analysis of the Ki67-stained epidermal wholemounts exemplified in (a), %; 
%REVISION
taken from a mouse aged 8 weeks; the dashed line shows the fit to the analytical form of the correlation function predicted by process (\ref{eqn:2DModel}), $C(x,t) = a(t) - \xi^{-1}(t) \ln x$ (see Eqs.~\ref{eqn:C(x,t)},~\ref{eqn:corrLength}), where the single time point $t$ is fixed by the age of the mouse and by initial conditions (see discussion in section~\ref{sec:disc}). From the fitted slope one may extract the correlation length $\xi(t)$ 
%
%the dashed line shows fitted slope used to extract the correlation length $\xi$ 
defined in Eq.~(\ref{eqn:corrLength}), giving $\xi=14.7$ cell diameters.
}
\label{fig:Ki67}
\end{center}
\end{figure}
As mentioned earlier, an intriguing feature of the non-interacting lattice model is the prediction of clustering of proliferating cells. How does this prediction compare with experiment? 
Before we turn to consider new results for mouse tail-skin, it is interesting to first consider
results presented in past work. In a detailed study of proliferating cells in Hamster cheek epidermis by Gibbs and Casarett~\cite{Gibbs:70}, a subset of progenitor cells were labelled using a radioactive marker for DNA synthesis (S-phase). By measuring the number of unlabeled cells separating consecutive labeled ones in one-dimensional basal layer cross-sections, it was possible to access the full radial distribution for the separation between adjacent cells undergoing S-phase. Remarkably, while the large-interval distribution decayed exponentially as expected for an uncorrelated random distribution, the probability of finding a nearby cell in S-phase was significantly higher at short distances, indicating that proliferating cells were clustered. A second study addressing the distribution of S-phase cells in mouse esophagus also revealed identical qualitative results~\cite{Cameron:66}. 

Not surprisingly, in the absence of the intuition afforded by the voter model, such clustering has been interpreted in the biological community as evidence of an underlying regulatory process,  which leads to the synchronous division of nearby cells~\cite{Cameron:66}. On the other hand, an independent analysis of clone fate data in Ref.~\cite{Klein:07} indicates that cell division within clones occurs independently. It is therefore satisfying to note that the tendency of proliferating cells to cluster \emph{is} in fact consistent with independent division --- indeed it is the hallmark of voter-model dynamics.

To extract the spatial correlation between progenitor cells in mouse tail skin, we analysed confocal micrographs of basal layer cross-sections of IFE that were stained for the proliferation marker Ki67, such as shown in Fig.~\ref{fig:Ki67}(a). Cells bright in Ki67 are designated as type A (progenitor) cells, whereas Ki67-dull cells were designated as type B (i.e. differentiated) cells, with no capacity to divide.
Using image analysis software (ImageJ), the coordinates of each Ki67-bright cell were extracted. This data allows a full statistical analysis of the spatial distribution of progenitor cells.  In particular, we shall focus on the radial correlation function
\begin{equation}
\label{eqn:corrFn}
C(x) = \left\langle \frac{1}{A} \int_0^{2\pi} \frac{d\theta}{2\pi} \int_A d\mathbf{x'}\, n(\mathbf{x'})\, n(\mathbf{x' +x}(x,\theta)) \right\rangle
\end{equation}
where $A$ denotes the area of each sample, $n(\mathbf{x})$ denotes the areal density of proliferating cells at position $\mathbf{x}$, and the brackets $\langle \cdot \rangle$ indicate averaging over all basal layer samples. 

Using Eq.~(\ref{eqn:corrFn}), the aim of the correlation analysis is to assess whether basal layer progenitor cells in adult mice do indeed cluster, and if so, to assess whether the experimental  correlation function is consistent with the predictions made in section~\ref{sec:voter}. In principle, one may look to the data for signatures of the expected spatial dependence $C(x,t)\sim a(t)-b(t)\ln x$ (for a given value of $t$), as well as for evidence of increased clustering with time, viz. $c_{\rm AB}(t)\sim 1/(2r\ln t)$. For the latter, however, the temporal analysis is difficult to implement due to the sensitivity required to resolve the $1/\ln t$ decrease in $c_{\rm AB}$ at long times. In particular, the predicted increase in clustering over a biologically-relevant period of $\lambda t\sim10^1-10^2$ cell cycles corresponds to a decrease in $c_{\rm AB}$ of $<10\%$, whereas variations in the efficiency of Ki67 labelling between different mice introduce systematic errors of the same order. Therefore, in the following we shall restrict ourselves to the quantitative analysis of tissue samples taken from a single adult mouse (aged 8 weeks).

To incorporate the empirical coordinates of proliferating cells into Eq.~(\ref{eqn:corrFn}), we replace the product $n(\mathbf{x}')n(\mathbf{x' +x})$ with the sum $\sum_{i,\, j} \delta(\mathbf{x'-R}_i) f(\mathbf{x'+x-R}_j)$, where $\delta(\mathbf{x})$ is the Dirac delta function, $\{\mathbf{R}_i\}$ is the set of all progenitor cell coordinates, and $f(\mathbf{X})$ is a two-dimensional Gaussian envelope with width $w$. With this substitution, the integrals in Eq.~(\ref{eqn:corrFn}) can be solved exactly. Then, explicitly accounting for the averaging procedure over sample of variable size, we obtain the expression
\begin{eqnarray}
\label{eqn:corrFnExp}
C(x) &=& \left\langle \frac{ \rho}{2\pi w^2 N_{\rm A}(x)} \sum_{i=1}^{N_{\rm A}} \sum_{j\ne i}^{N_{\rm A}(x)} \exp\left( -\frac{x^2 + R_{ij}^2}{2w^2}\right) \right. \nonumber \\
&&\ \ \ \ \ \ \ \ \ \ \ \ \ \ \ \ \ \ \ \ \ \ \ \ \ \ \ \ \ 
\times I_0\left( -\frac{x R_{ij}}{w^2}\right)\Bigg\rangle\, .
\end{eqnarray}
Here, the sum $i,\, j$ over all progenitor cell coordinates arises from the empirical expression for $n(\mathbf{x}')n(\mathbf{x' +x})$ given above, and we have defined $R_{ij} \equiv |\mathbf{R}_i-\mathbf{R}_j|$. The prefactor, exponential factor and modified Bessel function ($I_0$) result from solving the integrals in Eq.~(\ref{eqn:corrFn}). The total number of progenitor cells found in each sample is $N_{\rm A}$, and $N_{\rm A}(x)$ is the number of progenitor cells at a distance $x$ or more from the sample edges. As defined, the correlation function avoids errors resulting from edge effects by averaging each sample over the $N_{\rm A}(x)$ progenitor cells that are unaffected by the finite sample size. To correctly average over different samples, as indicated by $\langle \cdot \rangle$, the sample results are weighted by $N_A(x)$, e.g. ${C}^{(1+2)}(x) = [N_A^{(1)}(x) {C}^{(1)}(x) + N_A^{(2)}(x) {C}^{(2)}(x)]/[N_A^{(1)}(x)+ N_A^{(2)}(x)]$.

Making use of Eq.~(\ref{eqn:corrFnExp}), the experimental correlation between progenitor cells was averaged over 10 samples of approximately $30\times30$ cells each, see Fig.~\ref{fig:Ki67}(b). Remarkably, the data indeed shows a significant degree of progenitor cell clustering, in good agreement with the linear decay in $\ln x$ expected from Eq.~(\ref{eqn:C(x,t)}). In real terms, one may infer from the value of the nearest-neighbour correlation $C(1)$ that each progenitor cell is in contact, on average, with approximately \emph{two} adjacent progenitor cells, compared to $1.3$ progenitor cells expected for an uncorrelated random distribution (at $\rho=0.22$).

For a more careful test of the theory, one may extract from Fig.~\ref{fig:Ki67}(b) the order parameter $c_{\rm AB} = 0.29\pm0.02$, and the correlation length of $\xi = 14.7\pm 0.7$ cells, from which we find the empirical roughness constant, $\Omega = 0.04\pm0.01$. Repeating the analysis with samples taken from different mice results in the same value of $c_{\rm AB}$, but with values of $\Omega$ in the range $\Omega=0.02-0.04$.
Referring to Fig.~\ref{fig:voter}(b), where the empirical value of $\Omega$ is compared to the model predictions (dashed), we see that the model indeed recovers the correct order of magnitude of the roughness constant. Qualitatively, then, it appears that the model is consistent with the observed clustering. 
Let us emphasise that this fit requires \emph{no additional parameters}, and is purely a result of mapping the cell kinetics onto a lattice.

\section{Discussion and conclusions}
\label{sec:disc}
To summarise, we have demonstrated that the zero-dimensional model of cell division is consistent with the maintenance of the basal layer at uniform density, and we have
shown that the size distribution and qualitative shapes of labelled clones are consistent with a simple stochastic model of cell division and differentiation on a two-dimensional lattice. These results explain why the recently-discovered ``zero-dimensional'' model of cell behaviour gives such an excellent fit to the clone fate data despite taking no account of additional regulatory pathways. Significantly, despite the many forms of cell fate regulation known to exist in development and adult tissue, the \emph{only} extra-cellular regulation required to understand the existing observations of clone fate in normal IFE is \emph{steric}, viz. the coupling of cell division to the local cell density. 

%We note that the proposed model implements the constraint of uniform density by introducing the concept of a hard-core lattice in which density fluctuations are represented by lattice vacancies. However, it is clear that this simplified description captures the behaviour of range of analogous models for which the rigid lattice/vacancy description is not required, but which nevertheless relate between the local cell density and the capacity for progenitor cells to divide, see for example~\cite{shraiman:07}. 

Beyond the success of the model in explaining the observed clone fate data, we have also identified that the degree of progenitor cell clustering, as measured by Ki67 staining, is in good qualitative agreement with the predictions of the spatial process. 
In particular, there are two features that allow us to characterise the spatial process. First, the empirical roughness constant $\Omega=0.04\pm0.01$ has the expected order of magnitude predicted by the model (Fig.~\ref{fig:voter}b), and second, the correlation function is in excellent agreement with the expected decay form $a(t)-b(t)\ln x$ (at fixed $t$), as seen in Fig.~\ref{fig:Ki67}.

%REVISION
Taken together, these results have two important implications for future investigations of epidermal cell fate regulation: First, by demonstrating that the zero-dimensional process (\ref{eqn:model_0d}) is indeed capable of maintaining a uniform total basal layer cell density, the spatial process consolidates the proposed stochastic model as a robust platform for investigating biochemical constituents in future work. For example, by over/under-expressing specific genes and then studying the resulting change in the empirical parameters ($r,\ \lambda,\ \rho$), one may attempt to identify the role of each constituent in regulating cell behaviour. Second, the new model introduces a set of spatial measures (such as the roughness constant $\Omega$) that yield further information in such investigations of the biochemistry, beyond that which may be obtained through empirical evaluation of the zero-dimensional parameters alone.

Beyond the qualitative features of the dynamics, one may ask whether there are any implications to the quantitative features of the correlation analysis. Unfortunately, as mentioned earlier, variations in the efficiency of Ki67 labelling prevent us from generating the comprehensive statistics necessary to accurately quantify the universal values of the system parameters. Nevertheless, taken at face value, it appears that the empirical value of $\Omega$ obtained from the sample analysed in section~\ref{sec:corr} is only consistent with process (\ref{eqn:2DModel}) when one imposes a branching ratio $r=0.19\pm0.04$, which differs significantly from the value of $r=0.08$ established through clonal analysis (see fig.~\ref{fig:voter}b). This value is further consistent with the observed AB interface concentration of $c_{\rm AB}=0.29$ and the correlation length of $\xi = 14.7$ cells, which correspond to a long evolution time of $\lambda t \approx 10^4-10^6$ for $r=0.08$, but a realistic biological time-scale of $\lambda t \approx 10^1-10^2$ for $r=0.19$. 

Although it is possible that the discrepancy in the inferred value of $r$ results from errors in progenitor cell classification as discussed in section~\ref{sec:corr}, the difference is large enough to call into question the reliability of the correlation analysis. Beyond the issue of progenitor cell labelling efficiency, there is also the more basic question of whether Ki67 is at all effective as a marker of progenitor cells. In particular, it is widely accepted that Ki67 is a marker of cell growth as well as proliferation, which may imply that differentiated cells remain Ki67-bright for some time after division, or that progenitor cells may fail to continuously express Ki67~\cite{Brown_Gatter:02, Scholzen_Gerdes:00}. However, assuming that Ki67 is indeed a faulty marker, it becomes difficult to explain why the analysis nevertheless results in an excellent fit to the `$a(t)-b(t)\ln x$' decay form of correlations (at fixed $t$). %Therefore we are compelled to accept the data at face value.

We may also challenge our assumption of using ``random initial conditions'' to model steady-state maintenance. If initial conditions are at all relevant, then the observed clustering may be a signature of development and growth processes that are no longer active in the adult system. 
%However, any developmental process giving rise to such clustering would be severely constrained to give rise to the observed $a-b\ln x$ decay form of correlations. 
To test the importance of initial conditions, we have compared the order parameter $c_{\rm AB}$ between adult mice aged $8$ and $60$ weeks (corresponding to yound and old adult mice). If clustering is inherent to initial conditions, then one would expect $c_{\rm AB}$ to grow over time as the memory of initial conditions erodes. On the other hand, as mentioned in section~\ref{sec:corr}, steady-state maintenance results in $<10\%$ change in the order parameter over this time period, which would be undetectable given the systematic errors in Ki67 labelling. Indeed, we find no change from $c_{\rm AB}=0.29\pm0.02$ at $60$ weeks, so there is no indication that the observed clustering is a signature of tissue development. 
%However, a valuable verification of the model could still be obtained by considering the onset of progenitor cell clustering in confluent \emph{in vitro} cultures, using an initially homogenous sample of adult keratinocytes to eliminate the effects of tissue development.  

%Therefore, accepting that the observed clustering is both reliable and relevant to ongoing maintenance, we are led to look for 
Finally, it is interesting to speculate whether additional forms of regulation may be capable of reconciling the higher value of $r=0.19$ found for the spatial process, with the lower value of $r=0.08$ found from the zero-dimensional analysis. We propose a simple revision of the model that is capable of reconciling the spatial and zero-dimensional data: Referring to process (\ref{eqn:2DModel}), if the two channels of asymmetric division in process (\ref{eqn:2DModel}) occur with different probabilities (viz. $P_{\rm A\oslash\rightarrow AB}\ne P_{\rm A\oslash\rightarrow BA}\ne 1/2-r$, and $P_{\rm A\oslash\rightarrow AB}+ P_{\rm A\oslash\rightarrow BA}=1-2r$), then the effective value of the branching ratio $r$ is effectively renormalised in the spatial process, while leaving it unchanged in zero-dimensions. To reconcile the two empirical values of $r$, it is simple to show that one requires $P_{\rm A\oslash\rightarrow AB} \simeq 7P_{\rm A\oslash\rightarrow BA} \simeq (1-2r)/8$. Thus, one might speculate that the increase in clustering is associated with a spatial asymmetry in daughter cell fate during asymmetric division.
%, resulting from a sensitivity of the dividing cell to the local density gradient, or perhaps due to the orientation of nearest-neighbour progenitor cells. 
It would be an interesting challenge to devise an experiment with which to test this asymmetry.

{\sc Acknowledgements:} 
D. P. Doup\'e, P. H. Jones, and B. D. Simons acknowledge the financial support of the UK Medical Research Council. 

%%%%%%%%%%%%%%%%%%%%%%%%%%%%%%%%%%%%%%%%%%%%
%%%%%%%%%%%%%%%%%%%%%%%%%%%%%%%%%%%%%%%%%%%%

\end{document}